%% file: documentation.tex
\documentclass[onecolumn]{LaTeXclass/cohere}
\usepackage{ragged2e}
\usepackage{etoc}
\etocsettagdepth{mtmain}{none}
\etocsettagdepth{mtappendix}{subsection}
\makeatletter
\providecommand\@pnumwidth{1.55em}
\providecommand\@tocrmarg{2.55em}
\providecommand\@dotsep{4.5}
\makeatother
\usepackage{url}
\usepackage{booktabs}
\usepackage{tabularx}
\usepackage{array}
\usepackage{graphicx}
\usepackage{wrapfig}
\usepackage{placeins}
\usepackage{titlesec}

\title{Scaling Inference-Time Compute with Teams of Coding Agents}

\newcommand{\jointfirstauthor}{\fa}

\newcommand{\coherework}{\textsuperscript{\textcolor{DarkBlue}{$\boldsymbol{\dagger}$}}}
\AfterEndPreamble{{\footnotetext{\coherework Work done while at Cohere.}}}

\author{
    name={Keenan Samway\jointfirstauthor},
    affiliation={Cohere Labs}
}
\author{
    name={Daniel D'souza\jointfirstauthor},
    affiliation={Cohere Labs}
}
\author{
    name={Omar Darwiche Domingues},
    affiliation={Cohere}
}
\author{
    name={Marzieh Fadaee},
    affiliation={Cohere Labs}
}
\author{
    name={Tan Yi-Chern\coherework},
    affiliation={Cohere}
}
\author{
    name={Nouha Dziri},
    affiliation={Cohere Labs}
}
\newcommand{\emailpart}[1]{\textcolor{DarkBlue}{\texttt{#1}}}
\corresponding[*]{\{\emailpart{danieldsouza}, \emailpart{nouha.dziri}\}\emailpart{@cohere.com}}

\abstract{
\justifying
Multi-agent systems promise a new axis for scaling inference: rather than having one agent think longer or sample more attempts, a task is split across many agents that work in parallel and coordinate. However, in practice, more agents can easily produce duplicate work, coordination failures, and higher costs. We therefore formulate team design as a compute allocation problem: starting from a solo agent, where should additional inference compute go to reach the reward-cost Pareto frontier? We build a minimal multi-agent harness from the ground up and trace this frontier by varying, one at a time, the workflow stage at which a lead delegates to its workers (planning, implementation, or verification), the models that form the team, and the number of workers. On long-horizon challenging software-engineering tasks (e.g., ProgramBench), teams outperform solo agents but cost three to seven times as much. With the right compute allocation, however, a heterogeneous 17-agent team comes within 1.4 points of a strong homogeneous 5-agent team at 70\% lower cost. This efficiency stems from three findings. First, planning is the most valuable use of workers: in a homogeneous team whose lead decides how many workers to spawn, a lead that delegates only planning to its workers, and then implements and verifies the code itself, matches the reward of a lead that delegates to workers throughout planning, implementation, and verification, at half the cost. Second, cheap models pay off as planners but not as implementers: a strong lead with cheap planners recovers 47.5\% of the reward gain of an all-strong team for only 13.9\% of its additional cost, whereas cheap implementers lower reward. Third, planners keep improving reward as the team grows, whereas implementers and verifiers plateau beyond four workers: additional implementers fragment the code into more files for the lead to integrate, and additional verifiers add little new information about the program. Together, our results offer a guide to spending an agent budget where it counts, given the task, cost, and latency constraints a user has.}

\usepackage{titlesec}
\usepackage{xcolor}

\titleformat{\paragraph}[runin]
{\normalfont\bfseries\color{DarkBlue}}
{}
{0pt}
{}

\titlespacing*{\paragraph}{0pt}{0pt}{1em}

\begin{document}
\etocdepthtag.toc{mtmain}
\section{Introduction}

Coding agents can now solve long-horizon software-engineering tasks that require navigating large repositories, editing code, invoking tools, and iteratively debugging their own solutions \citep{jimenez2024swebenchlanguagemodelsresolve, yang2024sweagentagentcomputerinterfacesenable,deepseekai2026deepseekv41flashpushinglimitskv,zai2026glm53,kimiteam2026kimik3openfrontier,openai2026gpt56,anthropic2026fable5}. The dominant way to make such an agent stronger at inference time is to scale its compute sequentially, by letting it reason or revise for longer \citep{jaech2024openai,guo2025deepseek,muennighoff2025s1simpletesttimescaling}, or in parallel, by sampling independent attempts and aggregating them \citep{wang2022self,brown2024largelanguagemonkeysscaling,snell2024scalingllmtesttimecompute}. Both are bounded by a single agent's trajectory. Longer traces increase latency and eventually exhaust the context window, and independent attempts repeat the work rather than dividing it. A team of agents offers a third axis: it divides a single attempt among agents that work on different parts of the task, whose work must then be consolidated into one solution. Team members can explore competing hypotheses, inspect different parts of a repository, implement separable changes, and independently verify a candidate solution. This provides a new test-time scaling strategy for when a single agent's context window or time budget becomes limiting.

Adding agents, however, does not automatically add useful work. More agents bring more compute, which can raise solution correctness, but in practice these gains can be lost to duplicated work, conflicting edits, and the cost of consolidating what each agent produces \citep{geng2026effectivestrategiesasynchronoussoftware,yang2026parallelismpaysoffcohesionaware}. Whether a team helps therefore depends less on how many agents it has than on how it spends its compute.
We make this question precise by treating team design as a compute allocation problem (Section~\ref{sec:formulation}). Starting from a solo agent, the next dollar of inference can go to a stronger lead, to stronger or cheaper workers, to more workers, or to a particular stage of the software-engineering workflow \citep{royce1987managing}: planning (\textbf{P}), implementation (\textbf{I}), or verification (\textbf{V}). Each choice moves the team to a different point in reward and cost. We ask which choices reach the reward-cost Pareto frontier, and which buy the most reward per added dollar.

\begin{figure}[ht]
    \centering
    \includegraphics[width=\textwidth]{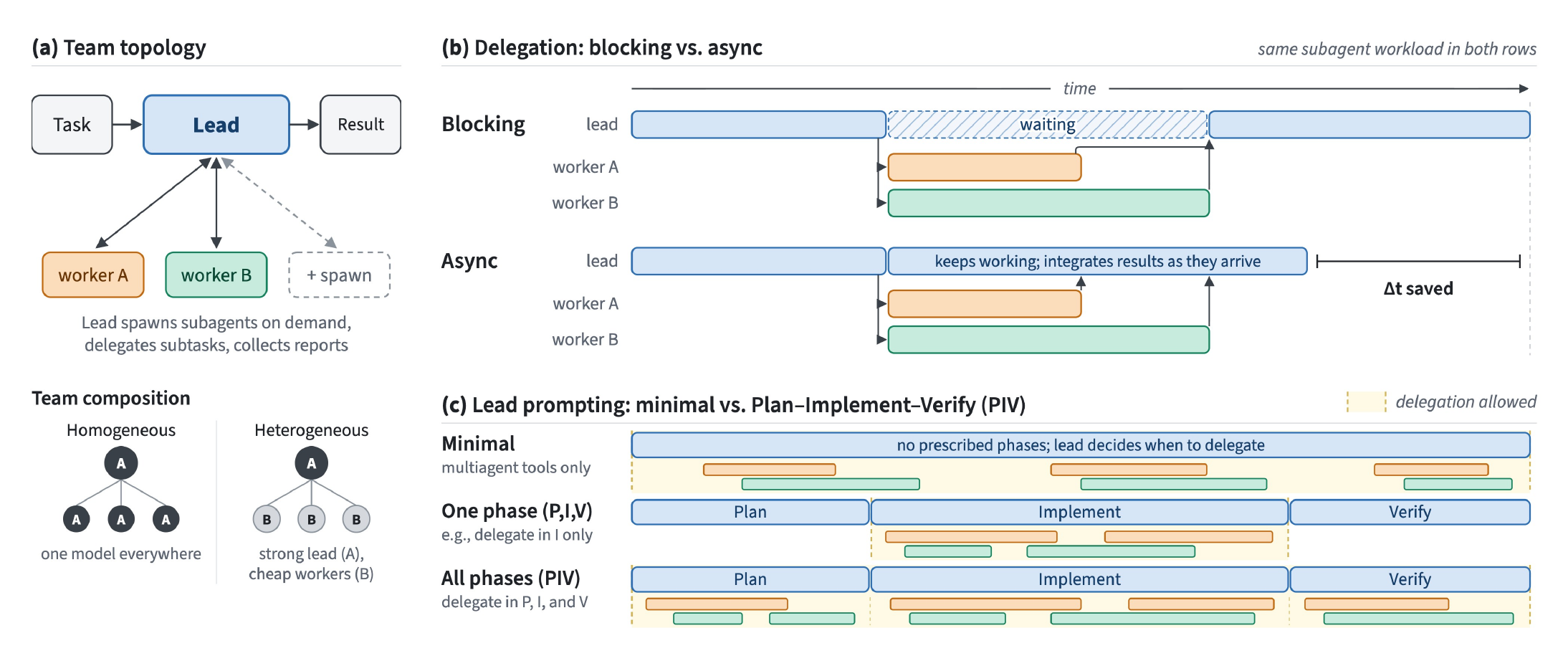}
    \caption{\small Overview of our multi-agent harness. A lead agent delegates subtasks to workers operating in a shared sandbox. We vary team composition (a), whether the lead blocks or keeps working while workers run (b), and which stages of the task the lead is prompted to delegate (c).}
    \label{fig:HARNESS}
\end{figure}

To answer this, we build a minimal, configurable harness with a hierarchical lead-worker topology (Figure~\ref{fig:HARNESS}): a lead agent decomposes the task, delegates subtasks to workers, and consolidates their results. Because every part of the harness is exposed, we can vary one design choice at a time: the stage at which the lead delegates (P, I, V, or all three, PIV), the lead and worker models, the number of workers, whether the lead waits for its workers or keeps working while they run, and whether workers can message each other. We focus on ProgramBench~\citep{yang2026programbenchlanguagemodelsrebuild}, a long-horizon repository-reconstruction benchmark in which agents receive only a black-box executable and its documentation, and must discover its behavior, plan, implement, and verify a complete program that reproduces it. We complement it with Commit0~\citep{zhao2024commit0librarygenerationscratch}, which also asks agents to build a repository from scratch but provides its skeleton, including the file layout and function stubs, together with its unit tests. We evaluate frontier models from the GPT and Claude Opus families.

Our results show that \textbf{teams outperform solo agents, but at a steep price}: a team composed of all strong agents raises reward over a single agent working alone, yet costs 3 to 7 times as much in inference dollars and, in most configurations, takes longer to finish (Section~\ref{sec:results-workers-guidance}). We find, however, that \textbf{with the right allocation of compute, a heterogeneous 17-agent team can come within 1.4 points of a strong homogeneous 5-agent team at 70\% lower cost} 
(Section~\ref{sec:scaling}). This configuration rests on where the compute goes. On ProgramBench, \textbf{planning emerges as the most valuable use of workers}: when the lead delegates planning to a team of workers and then writes and verifies the code itself, the team matches the reward of a lead that delegates all three stages (planning, implementation, and verification) to its workers, at \textbf{half its cost}, and each dollar spent on planning buys up to twice the reward of a dollar spent on any other stage (Section~\ref{sec:results-stage-guidance}). Adding more workers to implementation or verification, in contrast, quickly plateaus. Planning wins here because it is the bottleneck of this task; on Commit0, where the repository already dictates the plan, implementation benefits most instead (Appendix~\ref{app:commit0}). Because planning workers only gather information for the lead to act on, \textbf{they can be cheap}: a strong lead with cheap planners recovers 47.5\% of the reward gain of an all-strong team for only 13.9\% of its additional cost (Section~\ref{sec:heterogeneous}).

\section{Team Design as Budget-Constrained Optimization}
\label{sec:formulation}

A single agent spends its inference budget along one trajectory. A team, by contrast, distributes it across several design dimensions: the capability of the lead, the capability and number of workers, and the workflow stage at which workers are deployed. These dimensions are not independent, since the value of cheaper workers depends on the lead that directs them, and the value of additional workers depends on the stage to which they are assigned. We therefore formulate team design as a budget allocation problem.

\paragraph{Formulation} We consider hierarchical teams in which a lead agent decomposes the task, delegates subtasks to workers, and consolidates their results (Section~\ref{sec:experimental-setup}). A team configuration $c = (s, w, n, \ell)$ is specified by the stage $s \in \{\text{P}, \text{I}, \text{V}, \text{PIV}\}$ toward which the lead is instructed to direct its workers, the worker model $w$, the number of workers $n$, and the lead model $\ell$. On a task distribution $\mathcal{T}$, each configuration attains a mean reward $R(c;\mathcal{T})$ at a total inference cost $C(c;\mathcal{T})$.\footnote{We treat wall-clock latency as a secondary objective rather than a constraint in Eq.~\ref{eq:team-opt}. It is governed mainly by the scheduling policy (blocking or asynchronous), and we report it alongside reward and cost in Section~\ref{sec:results-workers-guidance}.} Ideally, designing a team under a budget $B$ would amount to solving
\begin{equation}
    R^*(B;\mathcal{T}) \;=\; \max_{c}\; R(c;\mathcal{T}) \quad \text{subject to} \quad C(c;\mathcal{T}) \le B,
    \label{eq:team-opt}
\end{equation}

whose solution, traced across budgets, defines the Pareto frontier of performance and cost. The optimal configuration is task-dependent: extra workers pay off most at the stage that bottlenecks the task (Section~\ref{sec:results-stage-guidance}, Appendix~\ref{app:commit0}).

\paragraph{Efficiency}
Solving Eq.~\ref{eq:team-opt} exactly is infeasible, as each configuration requires a full benchmark evaluation, and the stage instruction $s$ steers where the lead delegates rather than fixing how much it spends. We therefore do not match budgets across configurations. Instead, we ask a marginal question: \emph{starting from a solo agent, where does the next dollar of inference buy the most reward?} We define the \emph{efficiency} of a configuration $c$ as the reward it gains per dollar of added inference cost relative to the solo agent:
\begin{equation}
    \eta_c = \frac{R_c - R_{\text{solo}}}{C_c - C_{\text{solo}}},
    \label{eq:efficiency}
\end{equation}
where $R_c$ and $C_c$ denote the mean reward and total inference cost of configuration $c$, and $R_{\text{solo}}$ and $C_{\text{solo}}$ those of the solo agent under the same guidance. We report $\eta_c$ in reward points per \$100; a higher $\eta_c$ means the configuration spends its added compute more productively, and a negative $\eta_c$ means added compute lowers reward. A configuration \emph{dominates} another if it achieves at least the same reward at no greater cost, with one of the two strictly better. We compute $\eta_c$ only when $C_c > C_{\text{solo}}$; a configuration that costs no more than the solo agent while achieving higher reward dominates it, and we mark it as such.

\vspace{-2mm}
\paragraph{Tracing the frontier} We approximate the frontier by varying one design choice at a time and fixing each subsequent study at the most efficient choice found so far. We vary, in turn, the workflow stage $s$ at which the lead delegates to workers, whether planning, implementation, verification, or all three (Section~\ref{sec:results-stage-guidance}); the lead and worker models $\ell$ and $w$ (Section~\ref{sec:heterogeneous}); and the number of workers $n$ (Section~\ref{sec:scaling}). Since this coordinate-wise search does not solve Eq.~\ref{eq:team-opt}, the resulting configurations are empirical operating points rather than guaranteed optima.

\section{Experimental Setup}
\label{sec:experimental-setup}

\paragraph{Harness Architecture}
\label{sec:harness}

To study multi-agent orchestration under controlled conditions, we design a configurable harness that isolates key orchestration primitives such as scheduling, message delivery, agent prompts, and model choice.
Following the minimalist design of \texttt{mini-SWE-agent} \citep{yang2024sweagentagentcomputerinterfacesenable}, agents use a standard \texttt{bash} tool to interact with the environment and follow a ReAct-style loop \citep{yao2022react}.
We adopt a top-down hierarchy where a lead agent receives the task instruction and coordinates workers through a \texttt{delegate\_agent(task, agent\_id=None)} tool.
Omitting \texttt{agent\_id} instantiates a fresh worker with its own context window and role prompt.
Providing an existing ID resumes that worker, preserving its conversation history (e.g., to revise tests or continue debugging an authored module).
In some configurations, \texttt{list\_agents} and \texttt{wait\_for\_updates} tools are also provided to the lead, allowing it to check the status of all agents or block until the next worker update is surfaced.

\vspace{-2mm}
\paragraph{Agent Team Configurations} Given a task, the lead dynamically determines the team size and delegates well-scoped subtasks to worker agents. Each worker observes only its assigned subtask: it does not receive the full task description or observe other workers' conversations. We evaluate two scheduling strategies. Under \emph{blocking} scheduling, the lead waits for a worker to finish before continuing; under \emph{asynchronous} scheduling, the lead continues working while workers execute concurrently and receives their outputs upon completion. Unless otherwise noted, Sections~\ref{sec:results-workers-guidance}--\ref{sec:scaling} use only communication at delegation boundaries: the lead provides a subtask when it spawns or resumes a worker, and the worker's terminal output is delivered to the lead upon completion. Agents cannot exchange direct messages while a worker is running, either between the lead and workers or among workers. Section~\ref{sec:agent-communication} adds and evaluates this capability. We also compare two prompting regimes. Under \emph{minimal} prompting, the lead is only told to assign workers non-conflicting tasks and how its delegation tools behave; when and how to delegate is left to the lead. For guided prompting, we adopt a three-stage abstraction inspired by waterfall-style software development~\citep{royce1987managing}: planning and exploration (\textbf{P}), implementation (\textbf{I}), and verification (\textbf{V}). Each stage-specific guidance instructs the lead to complete the full workflow while using workers only in the designated phase. This design isolates the value of delegation at each phase. We then compose all three interventions into \textbf{PIV}, our \emph{heavy-guidance} variant, to evaluate their joint effect when workers are used throughout the workflow (full prompts in Appendix~\ref{app:prompts}).

\paragraph{Datasets}
We primarily use ProgramBench~\citep{yang2026programbenchlanguagemodelsrebuild}, a from-scratch repository reconstruction benchmark comprised of 200 unique tasks ranging from simpler command-line utilities like \texttt{CMatrix} and \texttt{Ripgrep} to complex programs like \texttt{FFmpeg} and \texttt{The PHP Interpreter}.
In each task, an agent is provided access to a fresh workspace, a black-box reference executable, and relevant behavioral documentation from the original repository.

Rather than repairing existing codebases or satisfying pre-written unit tests \citep{jimenez2024swebenchlanguagemodelsresolve}, the agent must actively probe the reference executable to discover its interface, edge cases, and runtime behavior, and then reconstruct a compatible implementation that is evaluated against held-out behavioral test suites.
Additionally, we also explore Commit0-Lite~\citep{zhao2024commit0librarygenerationscratch}\footnote{Commit0-Lite is only reported in the Appendix~\ref{app:commit0}: its 16 tasks limit statistical power, and it is near saturation, likely because frontier models are trained on its public libraries.} a benchmark comprised of 16 Python libraries that an agent is asked to reimplement given a repository skeleton and a set of unit tests. 

\paragraph{Models}

We conduct our main experiments with three tiers of the GPT-5.6 model family~\citep{openai2026gpt56}: Sol, Terra, and Luna, with all models set to use medium reasoning effort.
Homogeneous teams use the same model for the lead and all workers. Heterogeneous teams pair a Sol or Terra lead with Luna workers.
We also run additional experiments with Claude and Gemini model families (Appendix~\ref{app:additional-experimental-details}).

\section{Does a team of agents outperform a solo agent?}
\label{sec:results-workers-guidance}

\begin{figure}[ht]
    \centering
    \includegraphics[width=0.8\textwidth]{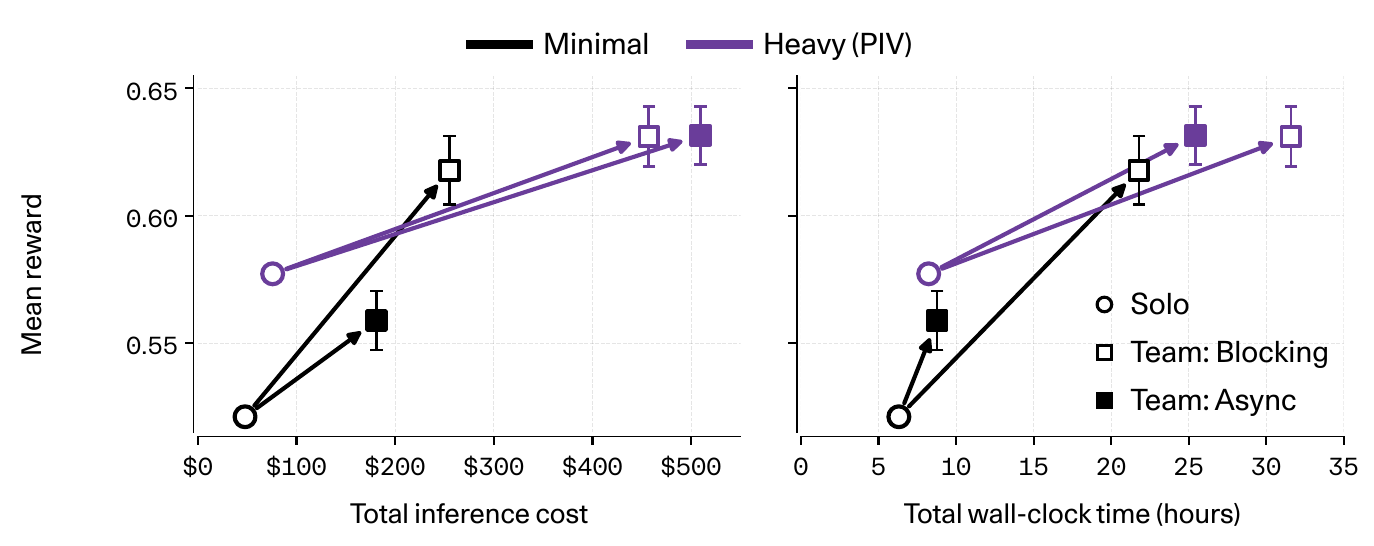}
    \caption{\small \textbf{Teams outperform solo agents, but incur higher inference and wall-clock costs.} Mean ProgramBench reward versus total inference
    cost (left) and total recorded wall-clock time (right) for Sol solo agents and homogeneous blocking and asynchronous teams under minimal and heavy (PIV) guidance. Points are means over three runs; error bars are 95\% intervals over tasks of each team's paired difference from the solo agent with the same guidance.
    \label{fig:programbench-solo-team-solo}}
\end{figure}

\paragraph{Results} 
Teams outperform solo agents under both minimal and heavy PIV guidance, and heavy guidance pushes them further ahead, yielding the best reward (Figure~\ref{fig:programbench-solo-team-solo}). Teams spawn between 4.2 and 5 workers depending on whether minimal or heavy guidance is provided. Compared to a solo agent that issues between 19 and 22 tool calls,  (Appendix~\ref{app:activity-stats}), teams issue between 44 and 140 tool calls per task based on the guidance respectively. Across configurations, gains are largest on easy tasks and are less pronounced on harder ones (Appendix~\ref{app:difficulty}). A team is thus an effective way to scale inference-time compute, but it incurs substantially higher inference cost. On Sol, blocking teams take 3.5$\times$ longer under minimal guidance and 3.8$\times$ longer under heavy guidance. Whereas asynchronous teams take 1.4$\times$ and 3.1$\times$ longer respectively, a setting where the lead can also progress simultaneously with the workers as opposed to blocking where the lead idles while the workers run. 
Under minimal guidance, however, this speed costs about 6 reward points relative to blocking. An asynchronous lead must decide for itself when to wait for workers, and in the trajectories we inspected it often acted on partial results or ended the task before its workers had finished. Heavy guidance, which instructs the lead to collect every report first, closes this gap.

This raises the central question of our study: 
\textit{how should a team allocate its compute to reach the Pareto frontier of performance and cost?}

\section{Where should workers be utilized?}

\label{sec:results-stage-guidance}


We first vary the stage $s$ to which workers are directed. We ask whether directing workers to specific stages of the software-engineering workflow (planning a design, implementing it, and verifying its behavior) changes this tradeoff. We consider 3 variants besides heavy guidance prompting (\textbf{PIV}) 
that differ in where the lead delegates work. Under \emph{heavy guidance (PIV)}, the team follows a plan-implement-verify workflow, and the lead decides which stages to delegate to workers (Appendix~\ref{app:prompts}). To isolate the value of each stage, in the other three variants we restrict all delegation to a single stage and let the lead handle the rest: under \emph{planning} (P), workers explore and plan while the lead implements and verifies; under \emph{implementation} (I), workers write code while the lead plans, integrates, and verifies; and under \emph{verification} (V), workers test the lead’s work after it plans and implements.
\footnote{We use Sol as the default model in subsequent analyses, since its higher capability better isolates how coordination choices shape team performance. Additional results for Terra and Luna are included in Appendix~\ref{app:additional-experimental-details}}


\begin{figure}[t]
    \centering
    \begin{minipage}[t]{0.48\textwidth}
        \centering
        \includegraphics[width=\linewidth]{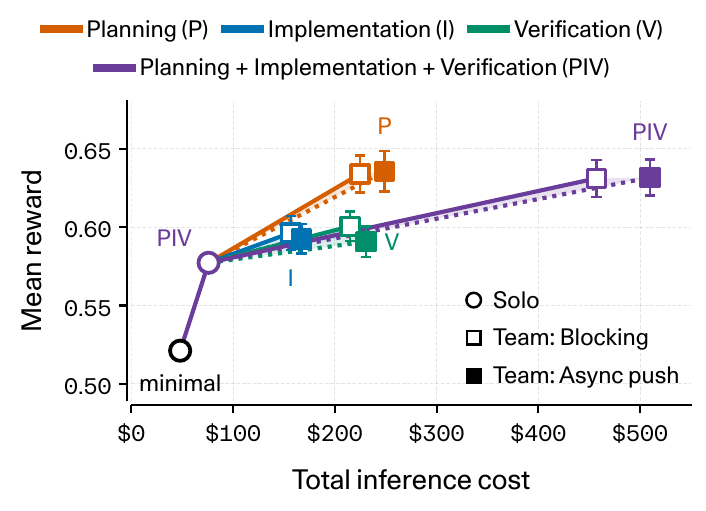}
        \caption{\small Planning provides the strongest reward--cost tradeoff under both schedules on ProgramBench (Sol teams; means over three runs).
    Error bars are paired 95\% intervals relative to Solo PIV.}
        \label{fig:programbench-stage-guidance}
    \end{minipage}\hfill
    \begin{minipage}[t]{0.48\textwidth}
        \centering
        \includegraphics[width=\linewidth]{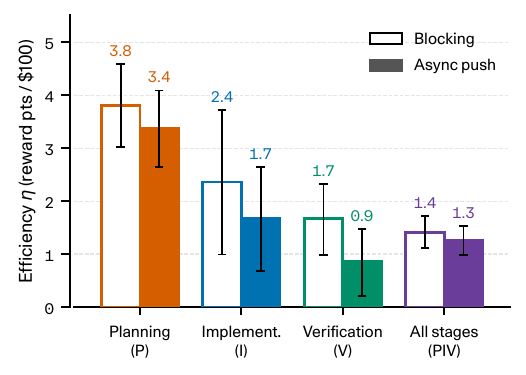}
        \caption{\small Efficiency $\eta$ of Sol teams relative to the Sol Solo PIV agent. Error bars are 95\% bootstrap intervals over tasks.}

        \label{fig:programbench-efficiency}
    \end{minipage}
\end{figure}

\paragraph{Results} Planning is by far the most efficient place to spend compute (Figure~\ref{fig:programbench-stage-guidance}, Figure~\ref{fig:programbench-efficiency}). With $\eta_{\text{P}} = 3.8$ versus at most $2.4$ for any other stage (Eq.~\ref{eq:efficiency}), each dollar spent on planning buys roughly $1.6\times$ more reward than the same dollar spent on implementation, verification, or the full workflow. Planning alone matches the reward of heavy PIV guidance at half the cost, and thus dominates it on the frontier. Because ProgramBench provides the target program only as a black-box executable, its behavior must be uncovered through probing, and planning workers parallelize this discovery across its interface, edge cases, and outputs. Additional compute therefore pays off most before any code is written, and delegating beyond planning adds cost without improving reward. 

\begin{figure*}[ht]
    \centering
    \includegraphics[width=\textwidth]{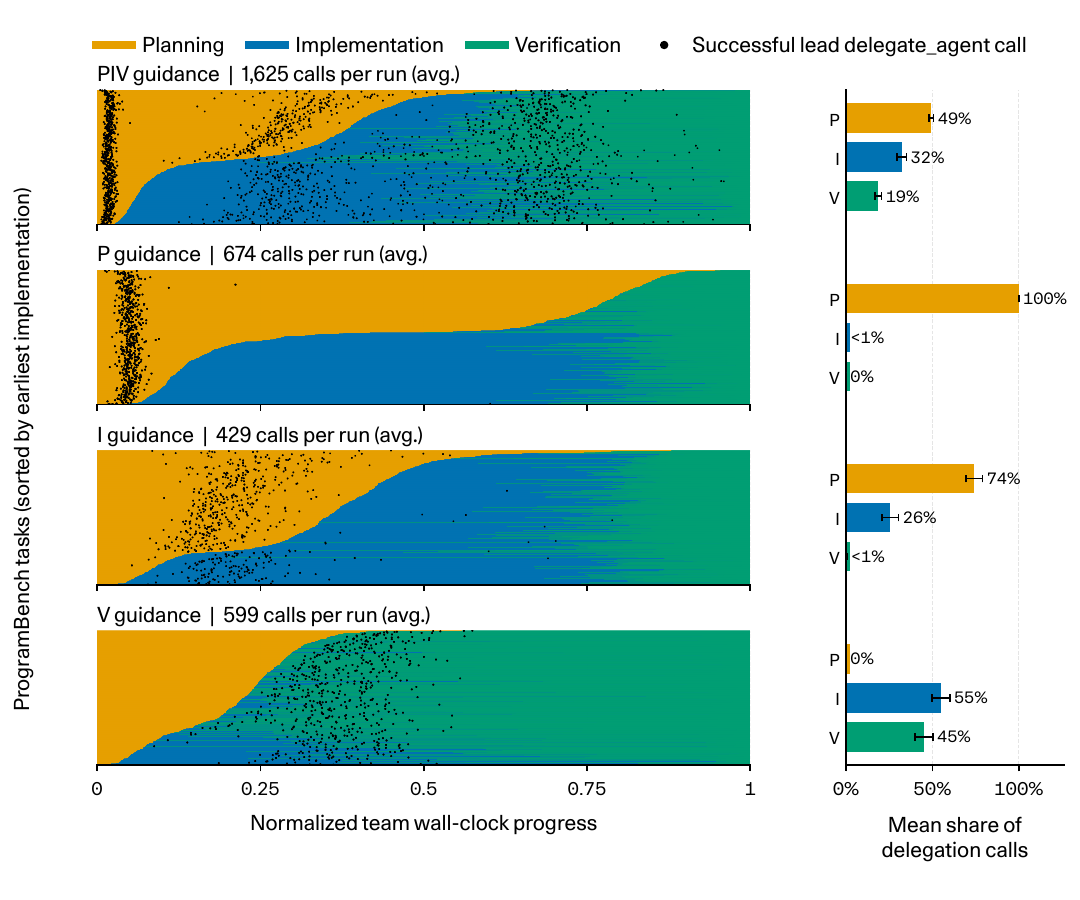}
    \caption{\small \textbf{Stage guidance shifts delegation timing, but nominal labels
    do not partition edit-defined phases.} The figure covers 600 ProgramBench rollouts per guidance setting (200 tasks $\times$ three runs) from homogeneous Sol blocking teams, with the lead's successful \texttt{delegate\_agent} calls marked. The four left rasters show PIV, P, I, and V guidance from top to bottom. Each row represents one rollout. Within each panel, rows are ordered by implementation start (the first detected source write; defined in Appendix~\ref{app:guidance-delegation}), with the earliest at the bottom; rows are not aligned across panels. The bars on the right give the mean share of a rollout's delegation calls in the planning (P), implementation (I), and verification (V) phases (one P-guidance rollout without delegation calls is excluded); error bars are 95\% bootstrap intervals from resampling tasks. Per-guidance call counts vary by at most 3\% across runs.}
    \label{fig:programbench-delegation-timing}
\end{figure*}

\paragraph{Why is planning efficient?} To understand how guidance shapes the team's use of compute, we locate every successful lead delegation call relative to the team's first and last source edits, which divide each rollout into planning (pre-edit), implementation (edit), and verification (post-edit) phases (Figure~\ref{fig:programbench-delegation-timing}). 

\newpage
On average, Planning guidance issues all of its 674 calls in a single burst before any code is written. Heavy PIV guidance issues a comparable 794 calls before the first edit, then keeps delegating through implementation and verification, adding an additional 830 calls that yield no additional reward. With the same gain over the solo agent, planning therefore obtains $2.4\times$ more reward per call, mirroring its efficiency advantage ($\eta_{\text{P}} = 3.8$ vs.\ $\eta_{\text{PIV}} = 1.4$): one early burst of discovery rather than continuous delegation.  This advantage is task-dependent. On Commit0-Lite, where the repository skeleton largely supplies the decomposition, implementation guidance instead becomes the most effective stage for delegation (Appendix~\ref{app:commit0}).


\begin{figure}[t]
    \centering
    \begin{minipage}[t]{0.48\textwidth}
        \centering
        \includegraphics[width=\linewidth]{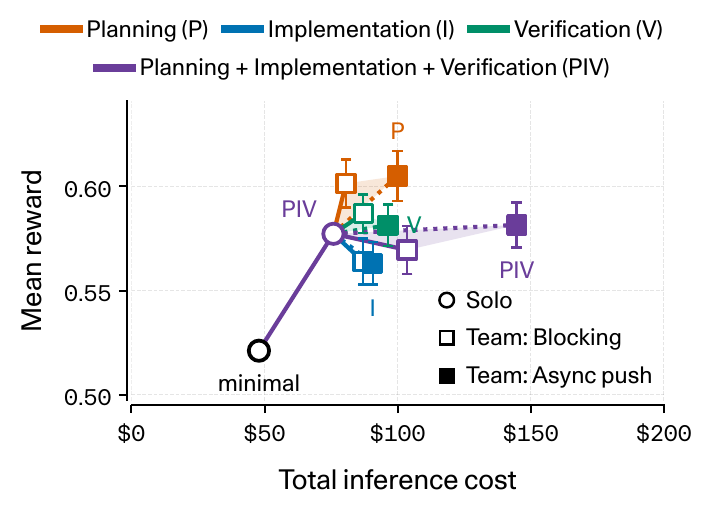}
        \caption{\small \textbf{Planning remains the strongest reward--cost tradeoff with cheaper workers.} Mean reward versus total inference cost for a Sol lead with Luna workers under P, I, V, or PIV guidance. Solid lines lead to blocking teams (open squares) and dotted lines to asynchronous teams (filled squares); error bars are paired 95\% intervals relative to Solo PIV.}
        \label{fig:programbench-sol-luna-stage-guidance}
    \end{minipage}\hfill
    \begin{minipage}[t]{0.48\textwidth}
        \centering
        \includegraphics[width=\linewidth]{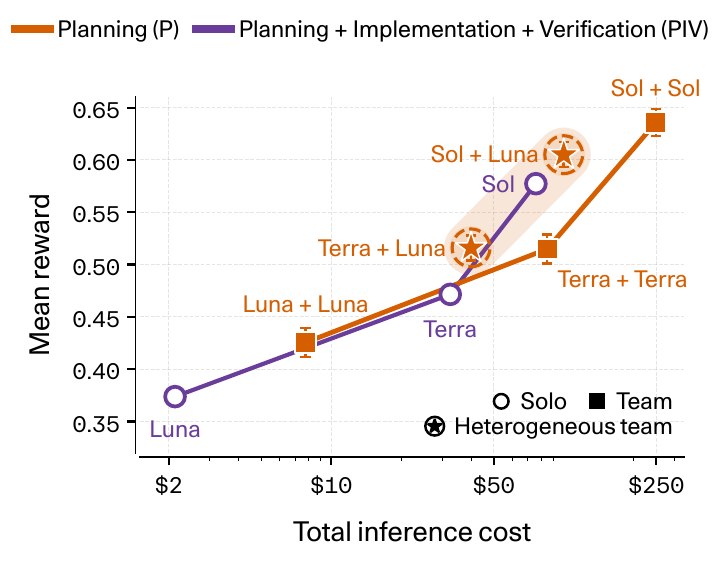}
        \caption{\small \textbf{Strong leads paired with cheaper planning workers open a lower-cost scaling axis.} Mean reward versus total inference cost (log scale) for Solo PIV agents and asynchronous planning teams led by Luna, Terra, or Sol. Stars mark heterogeneous teams with Luna workers: Sol+Luna recovers $47.5\%$ of the all-Sol team's reward gain at $13.9\%$ of its added cost.}

        \label{fig:programbench-heterogeneous-teams}
    \end{minipage}
\end{figure}

\section{Can Heterogeneous Teams Scale More Efficiently?}
\label{sec:heterogeneous}
In all configurations considered thus far, workers share the model of their lead. Replicating an expensive model across every role, however, multiplies the cost of the team, and it remains unclear whether strong workers are necessary to realize the benefits of delegation. To investigate this question, we retain a Sol lead and replace all of its workers with the substantially cheaper Luna. Because weaker workers may alter which stages of the workflow benefit from delegation, we again direct them toward planning (P), implementation (I), verification (V), or all three stages (PIV).

\paragraph{Results} Cheap workers pay off only when directed toward planning (Figure~\ref{fig:programbench-sol-luna-stage-guidance}). Relative to PIV, planning raises reward by 5.6\% while lowering cost by 22.2\% under blocking, and by 4.0\% while lowering cost by 30.9\% under asynchronous push. Directed toward implementation, by contrast, Luna workers reduce the team's reward below that of the Sol solo agent while increasing its cost, and directed toward verification, they leave reward essentially unchanged. The reward--cost trajectories in Figure~\ref{fig:programbench-sol-luna-stage-guidance} make this contrast explicit: planning climbs most sharply, whereas implementation falls below the solo level, where $\eta$ is negative. Cheap workers are thus effective when they gather information for a strong lead to act upon, but not when they produce code that the lead must integrate. The payoff of this substitution is striking: every dollar spent on Luna planners buys roughly $13\times$ more reward than the same dollar spent on Sol planners ($\eta = 51.0$ versus $3.8$ under blocking), whereas for implementation the same substitution is ruinous, with each additional dollar lowering reward ($\eta = -12.0$).

\paragraph{Can a stronger lead make better use of the same workers?} To disentangle lead capability from worker capability, we fix all workers at Luna and vary the lead across Luna, Terra, and Sol (Figure~\ref{fig:programbench-heterogeneous-teams}). Team reward closely tracks the capability of the lead: with identical Luna workers, upgrading the lead from Luna to Terra raises reward by 20.9\%, and from Terra to Sol by a further 15.4\%, and at every tier the team outperforms the solo agent built on the same model. Cheap workers, however, cannot compensate for a weak lead, as a Terra lead with Luna workers still trails a Sol agent working alone by 10.3\%. The lead therefore determines the ceiling that a team can reach, while the workers determine how cheaply it is approached, suggesting that model capability is best invested in the lead rather than replicated across the team. Across all lead tiers and schedules, planning remains the most efficient way to direct the workers, attaining the highest efficiency $\eta$ and reducing inference cost by 22--46\% relative to PIV (Appendix~\ref{app:lead-sweep}).

\section{How Many Workers Should a Team Have?}
\label{sec:scaling}

\begin{figure}[ht]
    \centering
    \includegraphics[width=0.9\textwidth]{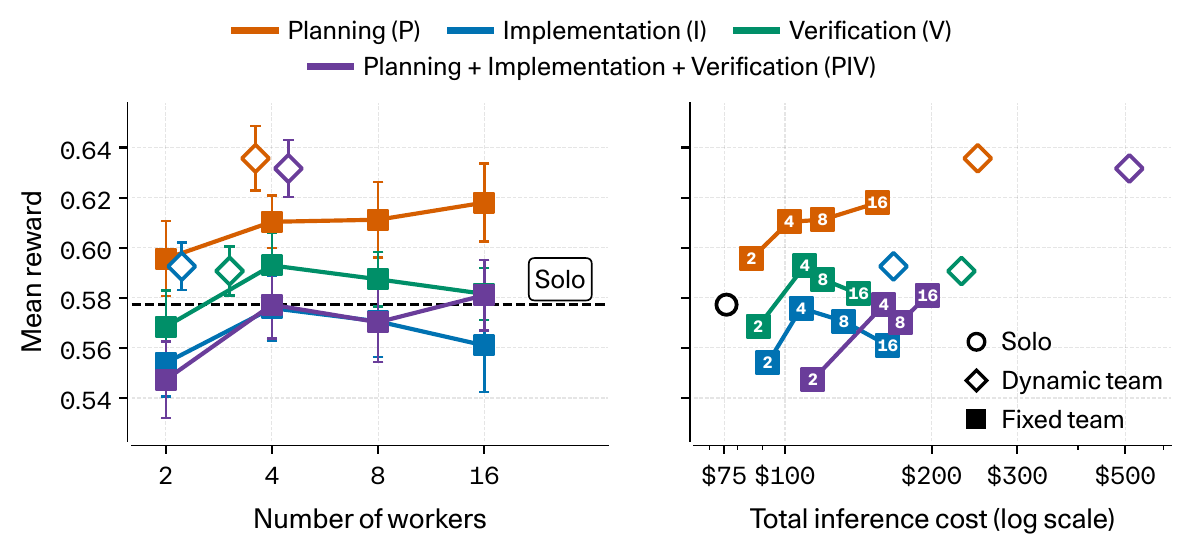}
    \caption{\small \textbf{Where additional agents are deployed determines the reward–cost scaling curve.} Fixed teams pair a GPT-5.6 Sol lead with 2--16 Luna workers (numbers give team size); dynamic teams are all-Sol teams sized by the lead, shown at their mean number of workers. Scaling the heterogeneous team to 16 workers with planning guidance comes within 1.4 points of the homogeneous team at 70\% lower cost. All teams use asynchronous push. Fixed teams are a single run; error bars (left) are paired 95\% intervals relative to Solo PIV.}
    \label{fig:agent-scaling}
\end{figure}

\paragraph{Scaling the number of workers} In the preceding sections, the lead determined the team size itself, spawning 4.2--5 workers on average. We now control the team size $n$ directly, pairing a Sol lead with 2, 4, 8, or 16 Luna workers directed toward planning, implementation, or verification, which lets us ask whether additional workers continue to pay off and at which stage of the workflow.
\textbf{Directed toward planning, sixteen Luna workers come within 1.4 points of an all-Sol team at 70\% lower cost}
(Figure~\ref{fig:agent-scaling}), making them roughly \textbf{$4\times$ more efficient}. This gain, however, depends on where the workers are directed, as Figure~\ref{fig:agent-scaling} reveals a pronounced asymmetry across stages. Planning workers improve reward at every team size: two workers already improve on the solo agent by 3.2\%, and sixteen raise the gain to 7.1\%.
Implementation and verification workers, in contrast, improve from two to four workers but then plateau, with their reward changing by $-0.015$ and $-0.011$, respectively, when scaling to 16 workers.
The plateau is not explained by explicit coordination failures.
We find that implementers rarely write to the same file (PIV guidance instructs the lead to delegate independent tasks; Appendix~\ref{app:prompts}) and verifier reports become less useful as the team grows (Appendix~\ref{app:scaling-failures}). Instead, planners return progressively more information about the black-box target, whereas additional implementers and verifiers subdivide work the lead can already coordinate without adding new information about the program.

\section{Does inter-agent communication help?}
\label{sec:agent-communication}

To study communication, we give every agent a \texttt{send\_message} tool that can reach any other running agent.\footnote{Agents were only able to send messages to other agents that were actively working, so we also provided all agents with the \texttt{list\_agents} tool.} Messages flow largely from workers to the lead (84\% for Sol teams under PIV guidance), carrying interim findings, proposed interfaces, verification failures, and questions. Workers rarely message each other (less than one message per task), even when told to collaborate (Appendix~\ref{app:peer-messaging}).

\begin{figure}[t]
    \centering
    \includegraphics[width=\textwidth]{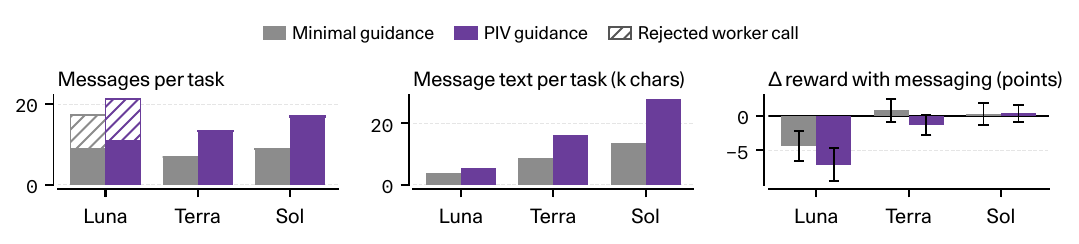}
    \caption{\textbf{Stronger models communicate more, but messaging does not raise reward.} Homogeneous asynchronous teams under minimal and PIV guidance. Hatched segments are rejected worker calls. Right: reward change relative to the same team without the tool, with paired 95\% intervals.}
    \label{fig:peer-messaging-main}
\end{figure}

Communication grows with model capability (Figure~\ref{fig:peer-messaging-main}). Under minimal guidance, all three tiers send 7--9 messages per task, but Sol writes $3.4\times$ as much text as Luna (13.7k versus 4.0k characters). Under PIV guidance, stronger teams also send more messages (10.9 to 16.9 per task), and their leads steer workers more (0.15 to 2.0 lead-to-worker messages per task). Greater communication, however, does not yield higher reward. Sol and Terra teams are unchanged within noise, and Luna teams lose 4.4--7.1 points (Figure~\ref{fig:peer-messaging-main}). The channel mostly carries status reports to a lead that already receives each worker's outcome.

\section{Related Work}
\label{sec:related-work}

\paragraph{Multi-agent Coding Systems} 
There is a growing body of work on teams of coding agents. Early systems encode a fixed team workflow into their orchestration, assigning specialized roles~\citep{qian2023chatdev,hong2023metagpt,tao2024magisllmbasedmultiagentframework,liu-etal-2025-rtadev}. More recent work asks how teams behave as they grow and run concurrently. CAID~\citep{geng2026effectivestrategiesasynchronoussoftware} combines dependency-aware delegation with isolated workspaces and asynchronous execution, and \citet{yang2026parallelismpaysoffcohesionaware} study when partitioning coding tasks across parallel agents pays off. Agensh~\citep{zhan2026agenshscalingorganizationalintelligence} lets teams self-organize from 1 to 128 agents on the five hardest ProgramBench tasks and scales one task to 1,024 agents. We instead hold one lead–worker architecture fixed and run a controlled study on all 200 ProgramBench tasks, measuring how each design choice moves the team along the reward-cost frontier.

 \paragraph{Team configuration under a budget} Closest to our work are studies of how to configure a team given limited compute. ZEBRA~\citep{hamri2026zebrazeroshotbudgetedresource} allocates a fixed monetary budget across coding phases, ManagerWorker~\citep{liu2026aimodelsdirectother} pairs strong managers with weaker workers, and AgentCARD~\citep{jiang2026specializerolesmixdeployments} shows that the value of role and model placement depends on the task domain. LoopArena~\citep{wang2026looparenabenchmarkingmodelsruntime} isolates controller capability by holding a coding worker fixed, and \citet{pan2026naturallanguageagentharnesses} and \citet{fan2026empiricalstudyharnessdesign} show that harness components such as planning, action interfaces, and context management change task outcomes and costs. We bring these questions together in a single white-box harness, comparing stage-directed delegation, scheduling, lead and worker capability, and team size on a common benchmark.

\section{Discussion and Conclusion}
\label{sec:discussion}

\paragraph{From Solo Agents to Agent Societies} Inference-time scaling is moving from single agents toward teams of agents, and we  expect this shift to accelerate as tasks grow from fixing a bug or implementing a feature to scientific discovery and open-ended research. Our results show that teams deliver gains over solo agents, but at a substantial increase in inference cost. We show, however, that with the right allocation of compute, a team can nearly match the reward of an all-strong team at 70\% lower cost. The next question is how to manage teams of hundreds, thousands, or even millions of agents. For this study, we considered teams of up to sixteen workers, although we expect team sizes to grow substantially in practice. At that scale,  how such teams should be organized? How should their agents communicate, and where is their compute best spent? And can agents optimize the network themselves, so that depending on the task, the budget, and the deadline, they make these choices naturally?
\paragraph{Imposed Hierarchy versus Self-Organization} Our harness imposes a lead–worker hierarchy, but a society of thousands of agents working over weeks must decide who assigns roles, how information flows, when to reorganize, and how agents communicate. In our runs, workers rarely talk to each other: 84\% of messages go to the lead and only 4\% between workers (Section~\ref{sec:agent-communication}). This may partly reflect our harness, which exposes a messaging tool but does not incentivize its use, and partly the task, since on ProgramBench the lead can still hold the full picture. As projects outgrow what any single agent can track, agents will need to share what they learn directly, and may even invent their own protocols to do so. Pretraining on human text already gives agents primitives such as reporting, asking, and delegating. Whether coordination emerges from these with model parameters scale or must be trained remains open, and we see it as central to scaling inference.
\paragraph{Call for Multi-Agent Safety and Security}
Our work studies how teams of coding agents scale and how to approach the reward-cost frontier, but does not evaluate their safety. The mechanisms that improve team performance may also introduce new risks: aligning individual agents does not guarantee that the team behaves as intended, errors can propagate through delegation, and agents may develop communication protocols that are difficult for humans to audit. We invite the broader research community, particularly those advancing multi-agent capabilities, to develop methods for monitoring, auditing, and aligning agent societies alongside methods for scaling them.

\section*{Acknowledgements}
We thank Beyza Ermiş, Julia Kreutzer and Jon Ander Campos for their thoughtful feedback on the paper; Gautier Dagan and Minjie Xu for early discussions; and Aryan Mann and Saurabh Dash for their technical assistance. We are also grateful to Thomas Euyang for the beautiful figures.

\nocite{*}
\bibliography{documentation}

\appendix
\etocdepthtag.toc{mtappendix}
\clearpage
\etocsettocstyle{{\noindent\LARGE\bfseries\color{DarkBlue}Appendices\par}}{} 
\tableofcontents

\section{Harness Architecture and Implementation Details}
\label{app:harness}

\subsection{Agent Turn Lifecycle}

A rollout begins by providing the benchmark task instruction as the initial user message to a lead agent.
The lead may invoke a \texttt{delegate\_agent} tool with a natural-language assignment to spawn and delegate a task to a subagent worker.
The new subagent initially begins with a fresh conversation state where the initial user message contains only this assignment from the lead.

When a subagent terminates, its final assistant message becomes the result returned to the lead.
Each worker is assigned a unique agent ID (e.g. \texttt{agent\_1}), and the lead can later delegate another task to a previously spawned agent, resuming the worker with its previous context and the new assignment as an appended user turn.
This persistence allows the lead to return to a worker that has already developed task-specific context, rather than starting a new trajectory for every delegation.

\subsection{Scheduling and Delivery Policy Comparison}

\begin{table*}[ht]
\centering
\caption{Comparison of scheduling and message-delivery policies.}
\label{tab:scheduling-delivery-policies}

\footnotesize
\setlength{\tabcolsep}{3pt}
\renewcommand{\arraystretch}{1.2}

\begin{tabularx}{\textwidth}{@{}
    >{\raggedright\arraybackslash}p{0.15\textwidth}
    >{\raggedright\arraybackslash}X
    >{\raggedright\arraybackslash}X
    >{\raggedright\arraybackslash}X
@{}}
\toprule
\textbf{Policy}
&
\textbf{How it works}
&
\textbf{Pros}
&
\textbf{Cons}
\\
\midrule

\textbf{Blocking delegation}
&
The lead delegates to a worker and remains blocked until the worker returns
its result.
&
Simple coordination; provides a natural synchronization point; allows the lead
to reason from a completed result; lowers the risk of acting on incomplete
information.
&
The lead is idle while the worker runs; prevents lead--worker parallelism; can
substantially increase wall-clock time.
\\

\midrule

\textbf{Async push: full message}
&
The worker runs concurrently. Once finished, its full result is appended to the
lead's next tool result.
&
Requires no explicit retrieval call; makes completed work difficult to
overlook; preserves concurrency; allows results to be incorporated quickly.
&
May interrupt the lead at an awkward point; unexpectedly consumes context;
multiple messages may overload the next tool output; may destabilize the
lead's current plan.
\\

\bottomrule
\end{tabularx}
\end{table*}

We study two scheduling policies. Under \textit{blocking delegation}, a call to \texttt{delegate\_agent} returns only when that worker has finished.
The lead therefore cannot begin its next step until the delegated work is completed.
However, multiple delegations issued in the same turn may execute in parallel, so blocking serializes the lead with respect to its workers, rather than serializing all activity.

Under \textit{asynchronous push}, a call to \texttt{delegate\_agent} immediately returns with the agent ID and the subagent continues to work in the background.
The lead can continue reasoning, using tools, or spawning additional workers while any subagents execute concurrently.
When a worker finishes, its outcome is appended to the subsequent tool result observed by the lead.
Optionally, the lead may call \texttt{wait\_for\_updates}, which blocks until a worker update becomes available.

\section{Guidance delegation}
\label{app:guidance-delegation}

We anchor each delegation call to the team's source
edits. The first and last detected source writes divide a rollout into three
observable intervals: time before editing, active editing, and time after
editing. Write boundaries use the
completion time of any lead or worker \texttt{bash} call matched by a
limited regex for patch commands, selected redirected shell writes,
terminal editors, or direct Python file writes. Writes under \texttt{/tmp},
\texttt{/var/tmp}, \texttt{/dev}, and \texttt{/opt}, as well as writes to
the generated \texttt{executable}, are excluded. 
 
We refer to these edit-defined intervals as planning,
implementation, and verification. They provide a common clock for comparing
trajectories across guidance ablations and studying delegation patterns.

Figure~\ref{fig:programbench-delegation-timing} applies this clock to every
successful lead delegation call from homogeneous Sol blocking teams on ProgramBench,
for PIV/P/I/V guidance setups. Calls rejected by the harness (e.g. exceeding the five-worker cap) are excluded from this analysis as they failed to successfully delegate a task to a worker.
We assign each call to an interval from its completion time relative to the edit boundaries, then
normalize that time by the rollout's team-wide wall-clock duration.

Planning guidance aligns most tightly with its nominal stage: all 674 P calls
occur before the first detected edit. V shifts calls later, although a slight
majority of its 599 calls remains within the active-edit window (55.0\%), with
45.0\% after the final edit. With I, 74.7\% of its 429 calls precede the first edit, 25.1\% occur
during active editing, and 0.2\% follow the last edit. PIV instead spans all
three phases, distributing its 1,625 calls across planning (48.9\%),
implementation (32.5\%), and verification (18.6\%). 

\section{Additional Experimental Details}
\label{app:additional-experimental-details}

All figures in this section use the full 200-task ProgramBench benchmark and the same conventions as the corresponding main-text figures. GPT-5.6 results are means over three runs; Claude results are single runs.

\subsection{Model configurations}
\label{app:model-configs}

Table~\ref{tab:model-configs} lists every model we evaluate, the provider we accessed it through, its reasoning setting, and the per-token prices from which all reported inference costs are computed. The same configuration is used whenever a model acts as a solo agent, a lead, or a worker.

\begin{table}[ht]
\centering
\caption{Model configurations and prices. Reasoning settings are OpenAI's reasoning effort, Anthropic's extended thinking (adaptive with an effort level, or a fixed token budget), and Gemini's thinking level. Prompt caching is enabled explicitly for Claude on Bedrock, which charges for cache writes; OpenAI and Gemini cache prompts automatically and charge only the discounted cache-read price. Evaluation gives the ProgramBench runs in which each model appears.}
\label{tab:model-configs}
\scriptsize
\setlength{\tabcolsep}{4pt}
\input{figures/appendix/model_configs_table.tex}
\end{table}

\subsection{Team versus solo across models}
\label{app:team-vs-solo-models}

\begin{figure}[ht]
    \centering
    \includegraphics[width=0.8\textwidth]{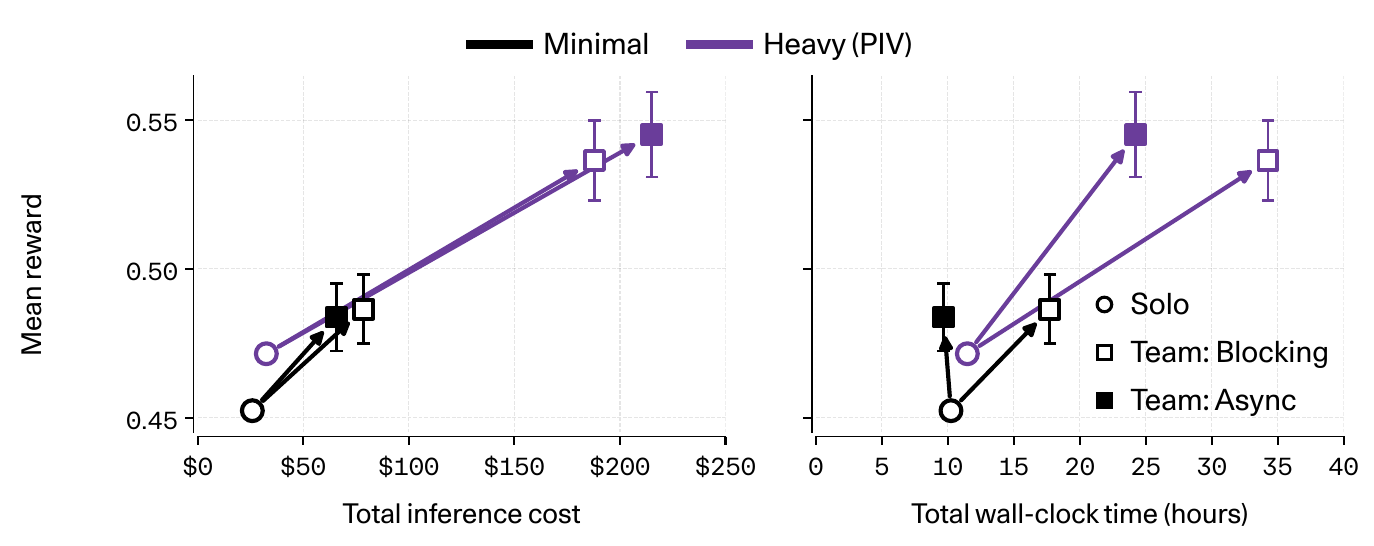}

    \caption{Solo agents versus homogeneous Terra teams. Each configuration is run for three trials, and error bars are 95\% intervals over tasks of each team's paired difference from the solo agent with the same guidance.}
    \label{fig:programbench-solo-team-solo-tiers-terra}
\end{figure}

\begin{figure}[ht]
    \centering

    \includegraphics[width=0.8\textwidth]{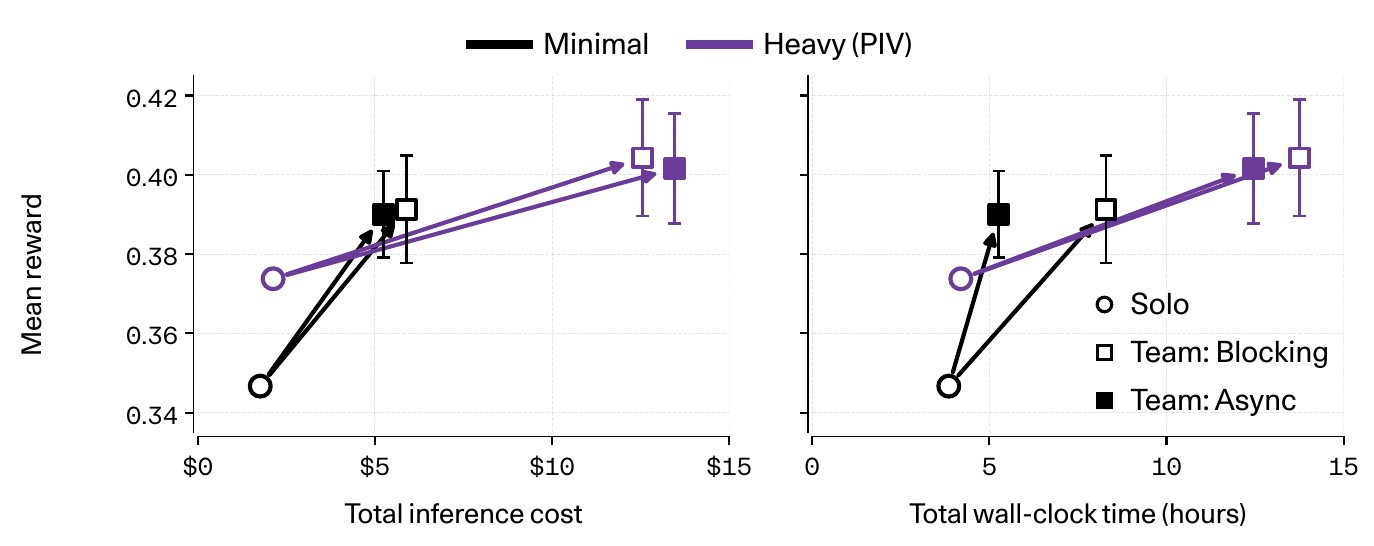}
    \caption{Solo agents versus homogeneous Luna (bottom) teams. Each configuration is run for three trials, and error bars are 95\% intervals over tasks of each team's paired difference from the solo agent with the same guidance.}
    \label{fig:programbench-solo-team-solo-tiers-luna}
\end{figure}

\begin{figure}[ht]
    \centering
    \includegraphics[width=0.8\textwidth]{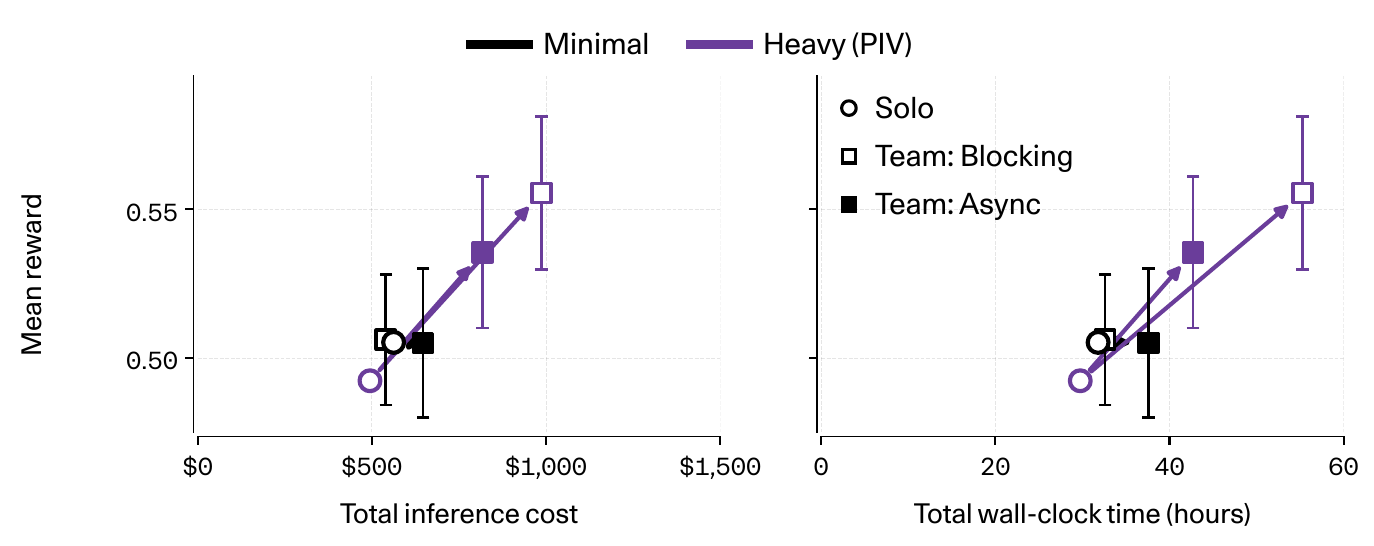}

    \caption{Solo agents versus homogeneous teams for Claude Opus 4.7. Each is a single run, and error bars are 95\% intervals over tasks of each team's paired difference from the solo agent with the same guidance.}
    \label{fig:claude-solo-team-opus}
\end{figure}

\begin{figure}[ht]
    \centering

    \includegraphics[width=0.8\textwidth]{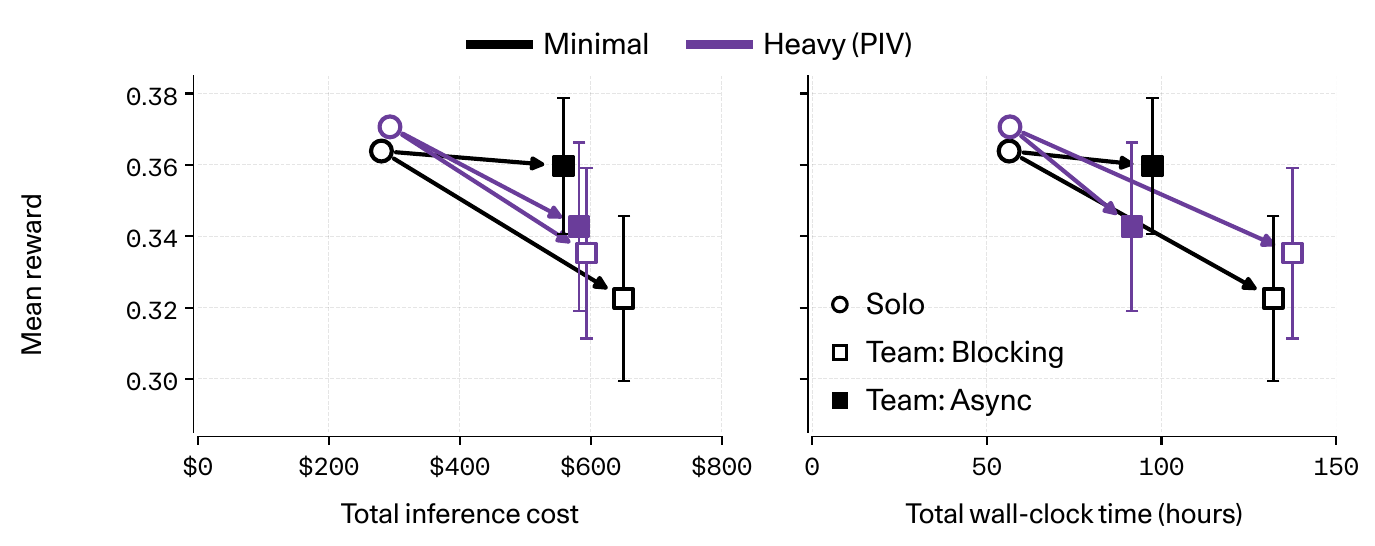}
    \caption{Solo agents versus homogeneous teams for Claude Haiku 4.5. Each is a single run, and error bars are 95\% intervals over tasks of each team's paired difference from the solo agent with the same guidance.}
    \label{fig:claude-solo-team-haiku}
\end{figure}

Here we repeat the solo-versus-team comparison for homogeneous GPT-5.6 Terra and Luna teams (Figures~\ref{fig:programbench-solo-team-solo-tiers-terra} and~\ref{fig:programbench-solo-team-solo-tiers-luna}) and for Claude Opus 4.7 and Haiku 4.5 (Figures~\ref{fig:claude-solo-team-opus} and~\ref{fig:claude-solo-team-haiku}). The Claude models illustrate two distinct ways in which added capacity fails to help. Opus 4.7 teams under minimal guidance are indistinguishable from the solo agent ($+0.001 \pm 0.022$), because the lead almost never delegates (Table~\ref{tab:delegation-stats-claude}). Under PIV guidance they gain $0.043$--$0.063$ at 1.7--2.0$\times$ the solo cost. Haiku 4.5 delegates readily, but every team configuration scores at or below the solo agent ($-0.041$ to $-0.004$) at roughly twice the cost.

\FloatBarrier
\subsection{Stage guidance across model tiers}
\label{app:stage-guidance-tiers}

\begin{figure}[ht!]
    \centering
    \begin{minipage}{0.49\textwidth}
        \centering
        \includegraphics[width=\linewidth]{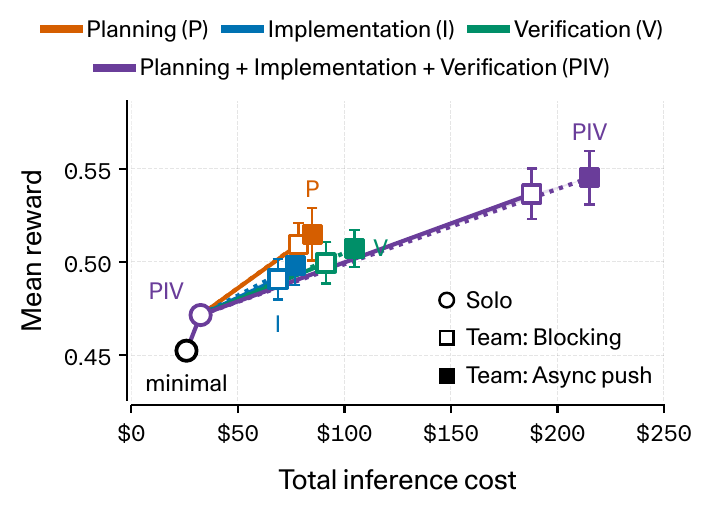}
    \end{minipage}\hfill
    \begin{minipage}{0.49\textwidth}
        \centering
        \includegraphics[width=\linewidth]{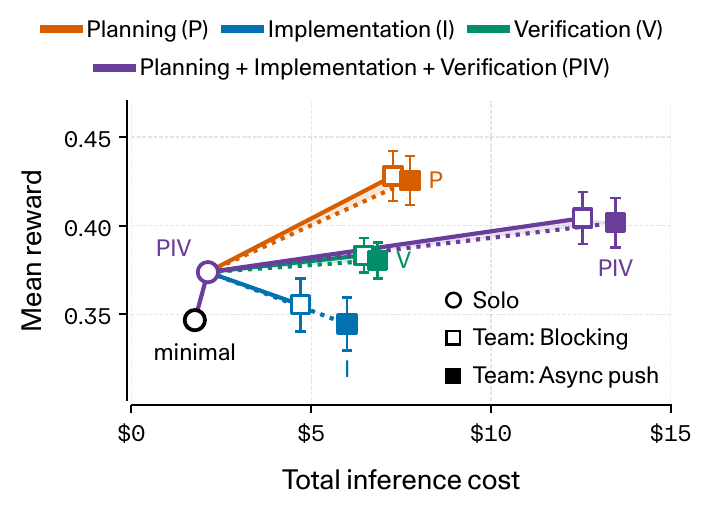}
    \end{minipage}
    \caption{Stage guidance for homogeneous Terra (left) and Luna (right) teams, as in Figure~\ref{fig:programbench-stage-guidance}.}
    \label{fig:stage-guidance-tiers}
\end{figure}

\begin{figure}[ht]
    \centering
    \begin{minipage}{0.45\textwidth}
        \centering
        \includegraphics[width=\linewidth]{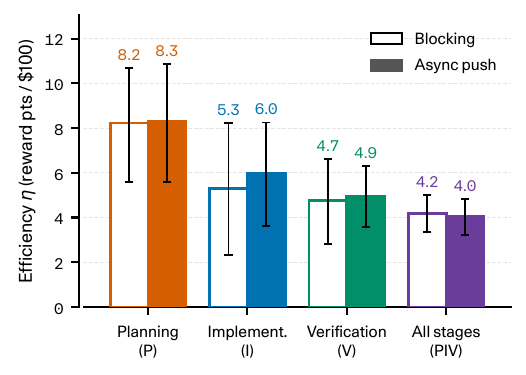}
    \end{minipage}\hfill
    \begin{minipage}{0.45\textwidth}
        \centering
        \includegraphics[width=\linewidth]{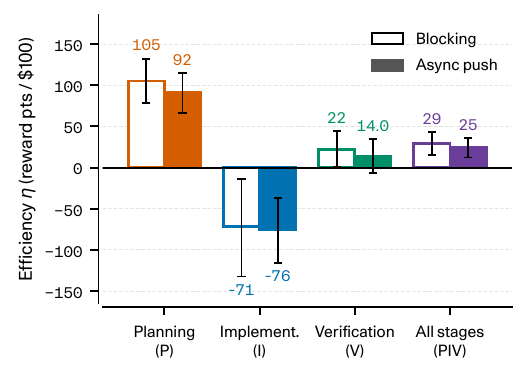}
    \end{minipage}
    \caption{Efficiency $\eta$ of homogeneous Terra (left) and Luna (right) teams relative to each tier's Solo PIV agent, as in Figure~\ref{fig:programbench-efficiency}.}
    \label{fig:efficiency-tiers}
\end{figure}

\begin{figure}[ht]
    \centering
    \includegraphics[width=0.5\textwidth]{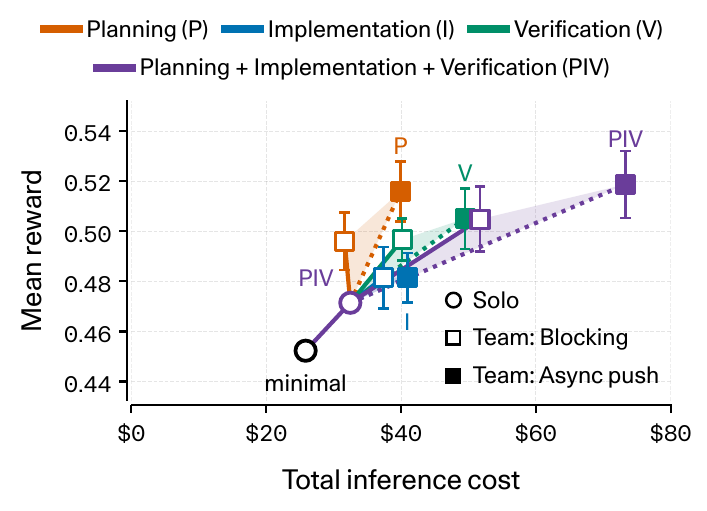}
    \caption{Stage guidance for heterogeneous teams with Terra lead with Luna workers.}
    \label{fig:terra-luna-stage-guidance}
\end{figure}

\begin{figure}[ht]
    \centering
    \begin{minipage}{0.45\textwidth}
        \centering
        \includegraphics[width=\linewidth]{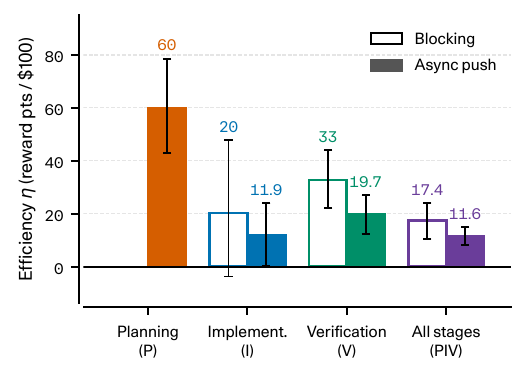}
    \end{minipage}\hfill
    \begin{minipage}{0.45\textwidth}
        \centering
        \includegraphics[width=\linewidth]{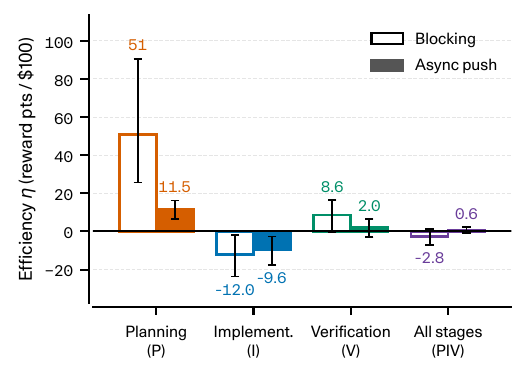}
    \end{minipage}
    \caption{Efficiency $\eta$ for a Terra lead (left) and a Sol lead (right) with Luna workers, relative to the lead's Solo PIV agent. A missing bar marks a team that is both cheaper and higher-reward than the solo agent (it dominates it). Error bars are 95\% bootstrap intervals over tasks.}
    \label{fig:efficiency-luna-workers}
\end{figure}

Figures~\ref{fig:stage-guidance-tiers} and~\ref{fig:efficiency-tiers} repeat the stage-guidance (Figure~\ref{fig:programbench-stage-guidance}) and efficiency (Figure~\ref{fig:programbench-efficiency}) comparisons for homogeneous Terra and Luna teams. Figures~\ref{fig:terra-luna-stage-guidance} and~\ref{fig:efficiency-luna-workers} do the same for teams with Luna workers, complementing Figure~\ref{fig:programbench-sol-luna-stage-guidance}.

\FloatBarrier
\subsection{Heterogeneous teams and lead capability}
\label{app:lead-sweep}

\begin{figure}[ht]
    \centering
    \includegraphics[width=0.48\textwidth]{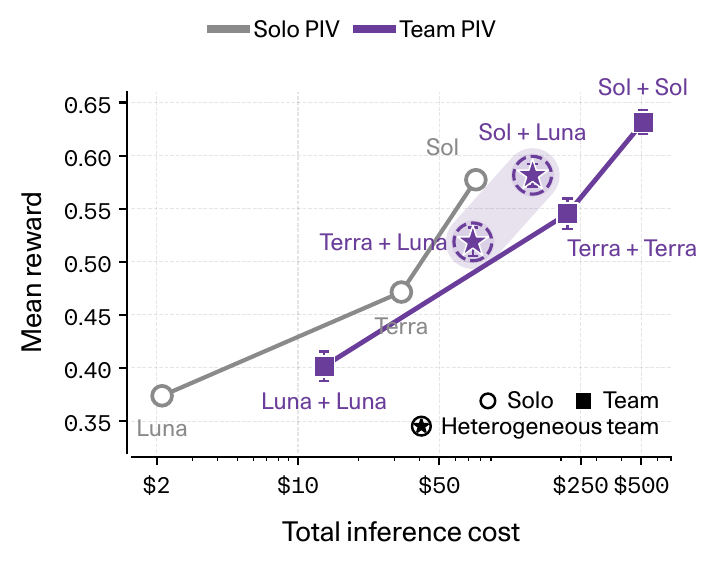}
    \caption{Solo PIV agents versus homogeneous and heterogeneous (Luna-worker) teams under PIV guidance, the PIV counterpart of Figure~\ref{fig:programbench-heterogeneous-teams}. Teams use asynchronous push.}
    \label{fig:heterogeneous-teams-piv}
\end{figure}

\begin{figure}[ht!]
    \centering
    \includegraphics[width=0.8\textwidth]{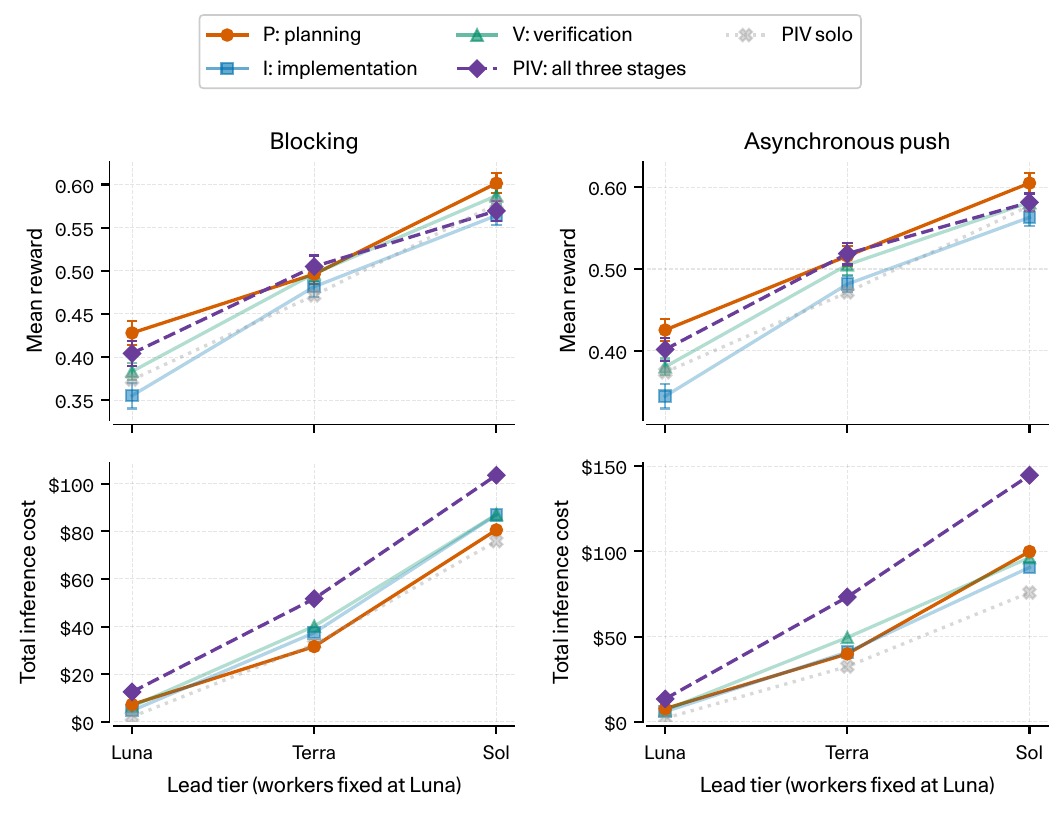}
    \caption{Mean reward (top) and total inference cost (bottom) with workers fixed at Luna and the lead varied across Luna, Terra, and Sol, under blocking (left) and asynchronous push (right). Faint lines show implementation and verification guidance; the gray dotted line is each lead's Solo PIV agent. Reward error bars are paired 95\% intervals relative to each lead's Solo PIV agent.}
    \label{fig:lead-sweep}
\end{figure}

Figure~\ref{fig:heterogeneous-teams-piv} repeats Figure~\ref{fig:programbench-heterogeneous-teams} with PIV-guided instead of planning-guided teams.
Figure~\ref{fig:lead-sweep} fixes workers at Luna and varies the lead. With workers fixed at Luna, planning guidance attains the highest reward with Luna and Sol leads under both schedules; with a Terra lead, PIV is 0.003--0.009 higher. Planning costs 22--46\% less than PIV across all six lead--schedule combinations (Figure~\ref{fig:lead-sweep}).

\FloatBarrier
\subsection{Run-to-run Variance}
\label{app:variance}

\begin{figure}[ht]
    \centering
    \includegraphics[width=0.85\textwidth]{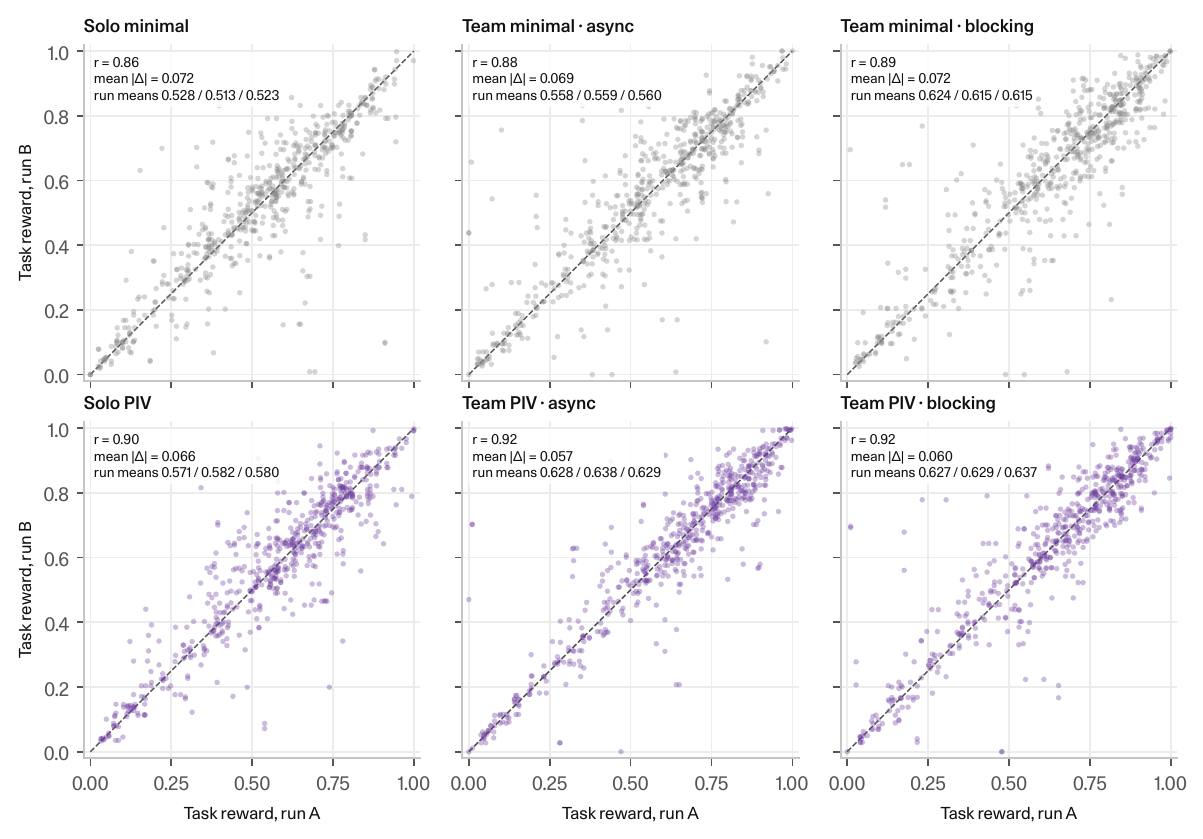}
    \caption{Task-level reward comparisons between two GPT-5.6 Sol solo agent and teams runs (top: minimal guidance; bottom: PIV guidance). Each panel pools three of runs (600 points total); the dashed diagonal line marks equal reward. Annotations give the mean pairwise correlation, the mean absolute difference, and each run's 200-task mean.}
    \label{fig:variance-sol}
\end{figure}

Our main results with the GPT 5.6 series (Sol, Terra, Luna) are run and averaged across three independent trials on the full set of 200 ProgramBench tasks.
Figure~\ref{fig:variance-sol} compares task-level rewards between pairs of runs for six Sol configurations.
Individual task outcomes are noisy; across all five model setups, the same configuration's reward on a task differs between two runs by 0.057--0.113 on average.
However, rewards remain strongly correlated across runs and benchmark-level means are much more stable (Table~\ref{tab:variance}).
Across the five model setups, the standard deviation of the 200-task mean across runs is at most 0.014.

\begin{table}[p]
\centering
\caption{Run-to-run variance for every GPT-5.6 configuration across three runs. All summary statistics are computed over 200-tasks across three runs. Task $|\Delta|$ is the mean absolute difference in a single task's reward between combinations of any two runs; task correlation is the mean pairwise correlation of task rewards between runs.}
\label{tab:variance}
\scriptsize
\setlength{\tabcolsep}{5pt}
\input{figures/appendix/variance_table.tex}
\end{table}

\subsection{Activity Statistics}
\label{app:activity-stats}

\begin{figure}[ht]
    \centering
    \includegraphics[width=\textwidth]{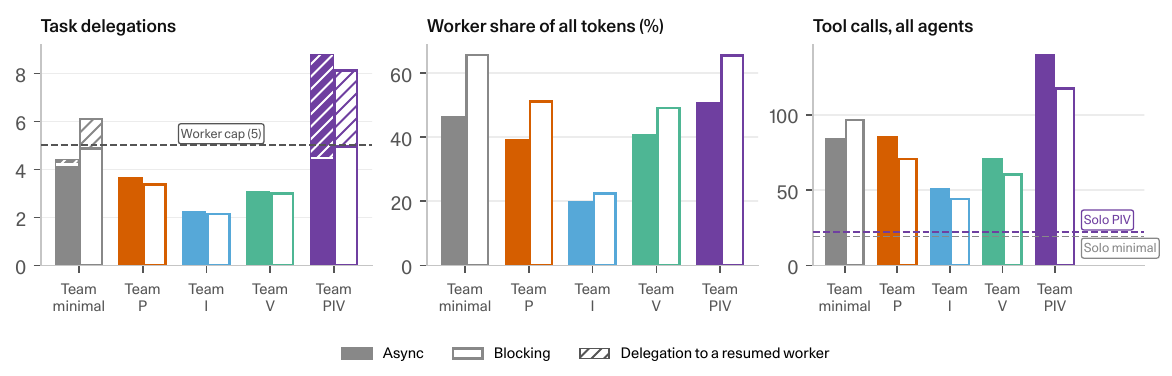}
    \caption{Per-task team activity for GPT-5.6 Sol teams pooled over three runs. Filled bars denote asynchronous and outlined bars blocking teams. In the delegation panel, the solid part of each bar counts delegations that spawn a new worker (and therefore equals the number of workers spawned), hatching marks delegations that resume an already spawned worker, and the dashed line marks the standard cap of five unique worker agents. In the tool-call panel, dashed lines mark Solo minimal and Solo PIV.}
    \label{fig:team-activity}
\end{figure}

Tables~\ref{tab:delegation-stats} and~\ref{tab:delegation-stats-claude} and Figure~\ref{fig:team-activity} summarize how each configuration uses its team.
A solo Sol agent issues 19--22 tool calls per task, whereas teams issue between 44 and 140.
Stage guidance largely determines team size: the lead spawns 2.2 workers on average under implementation guidance, 3.0 under verification, 3.4--3.6 under planning, and 4.2--5.0 under minimal and PIV guidance, where the cap of five unique workers binds.
Persistent workers are reused almost exclusively under PIV, where 38--48\% of delegations resume a worker that already holds task context.
Implementation guidance leaves the lead with most of the work, as workers account for only 21\% of tokens compared with 46--66\% under minimal and PIV guidance.

\begin{table}[p]
\centering
\caption{Per-task team activity for GPT-5.6 models, grouped by lead model and worker model. All columns are per-task means pooled over three runs. Cost is inference cost per task; the resumed share is the fraction of delegations that resume a previously spawned worker, and the worker token share is the fraction of all tokens generated by workers.}
\label{tab:delegation-stats}
\scriptsize
\setlength{\tabcolsep}{4pt}
\input{figures/appendix/delegation_stats_table_gpt.tex}
\end{table}

\begin{table}[ht]
\centering
\caption{Per-task team activity for Claude models on the full 200-task ProgramBench run (single run; same columns as Table~\ref{tab:delegation-stats}).}
\label{tab:delegation-stats-claude}
\scriptsize
\setlength{\tabcolsep}{4pt}
\input{figures/appendix/delegation_stats_table_claude.tex}
\end{table}

\FloatBarrier
\subsection{Where teams add reward: task difficulty}
\label{app:difficulty}

We compare each team with the solo agent that receives matching guidance: minimal teams with the minimally prompted solo agent, and stage-guided teams (P, I, V, and PIV) with the solo agent given the full plan--implement--verify prompt.
Across configurations, the lift generally decreases from easy to hard tasks.
The one exception is for Implementation guidance, where the trend reverses, and we observe the team generally performing worst on the easy tasks.

\begin{figure}[ht]
    \centering
    \includegraphics[width=0.85\textwidth]{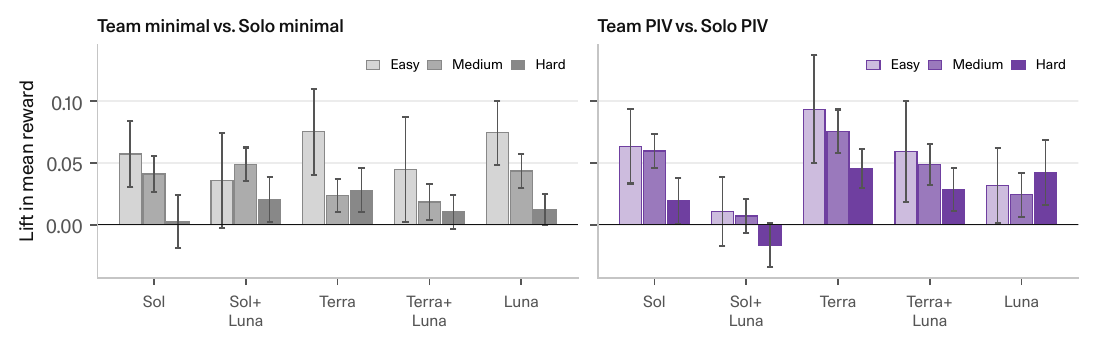}
    \caption{Lift by difficulty tier for asynchronous Team minimal (relative to Solo minimal) and Team PIV (relative to Solo PIV), for all model setups. Lighter shades (leftwards in each group) denote easier tiers; error bars are 95\% intervals over tasks.}
    \label{fig:difficulty-setups-core}
\end{figure}

\begin{figure}[ht]
    \centering
    \includegraphics[width=0.85\textwidth]{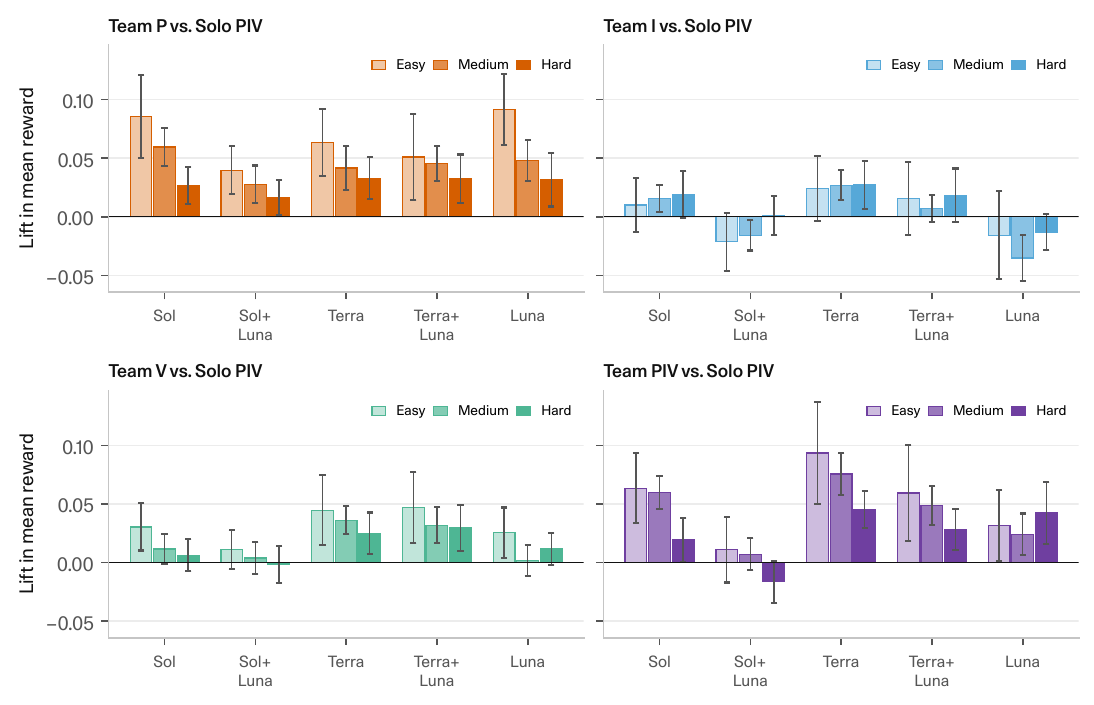}
    \caption{Lift by difficulty tier for asynchronous Team P, I, V, and PIV relative to Solo PIV, for all model setups.}
    \label{fig:difficulty-setups-stages}
\end{figure}

\FloatBarrier
\subsection{Why implementation and verification plateau}
\label{app:scaling-failures}

In the fixed-size runs (Figure~\ref{fig:agent-scaling}), planning is the only stage whose reward monotonically grows with team size ($+2.2$ points, 95\% CI $[+1.0, +3.5]$ from two to sixteen workers). Implementation and verification improve from two to four workers and then plateau. From four to sixteen workers they change by $-1.5$ points, 95\% CI $[-3.4, +0.4]$ and $-1.1$ points, 95\% CI $[-2.6, +0.3]$, respectively.

\begin{figure}[ht]
    \centering
    \includegraphics[width=\textwidth]{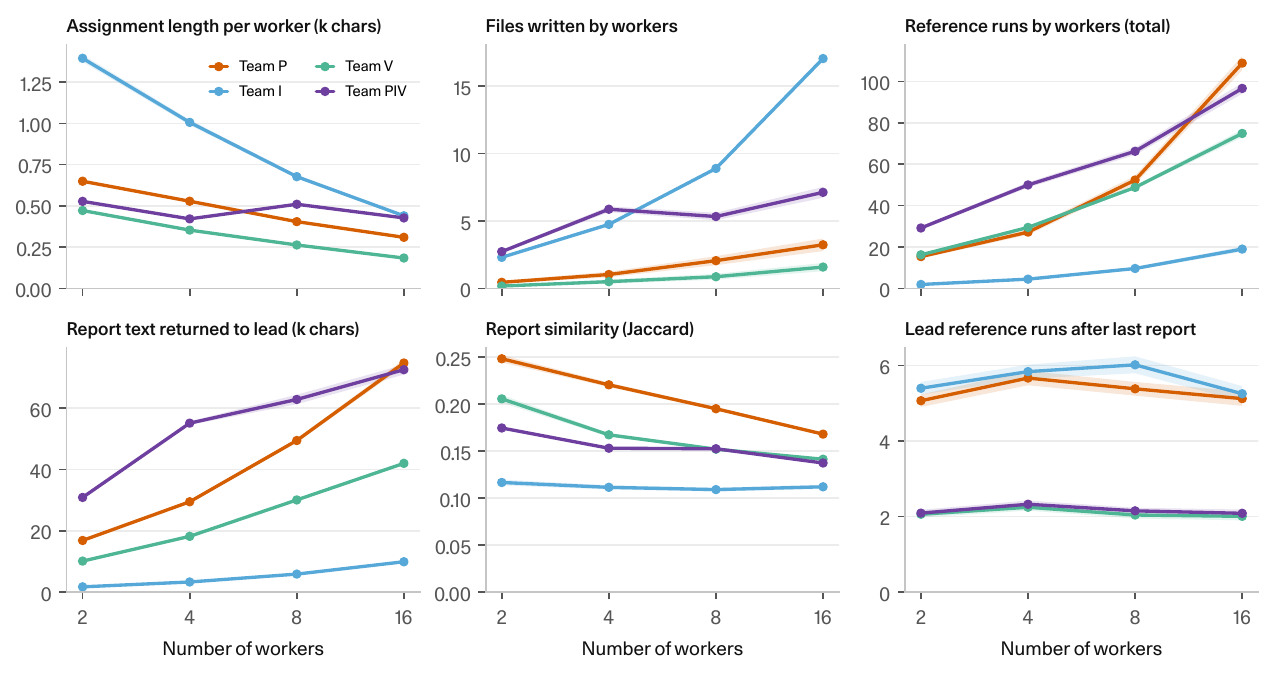}
    \caption{How work is divided as fixed-size teams grow (asynchronous Team P, I, V, and PIV; Sol lead with 2--16 Luna workers). Panels show the length of each worker's assignment, files written by workers, total runs of the reference executable by workers, total report text returned to the lead, pairwise report similarity (word-set Jaccard; higher means more alike), and the lead's runs of the reference executable after the last worker report. Files and reference runs are detected from shell commands, so file counts are lower bounds.}
    \label{fig:scaling-failures}
\end{figure}

Figure~\ref{fig:scaling-failures} shows how work is divided as teams grow. Explicit coordination failures are rare: implementers almost never write the same file (0.1\% of their files), and worker reports become less similar as the team grows. What differs is what an additional worker contributes. Each planner runs the reference executable about seven times regardless of team size, so the information returned to the lead grows with the team (15 to 109 reference runs and 17k to 75k characters of reports from two to sixteen workers), while the Sol lead still writes essentially all of the code. Adding implementers instead subdivides a fixed plan: the lead explores for the same number of turns before delegating, each worker's assignment shrinks from 1.4k to 0.4k characters, implementers barely consult the reference (about one run each), and the share of source edits made by the Luna workers rises from 54\% to 85\%. The lead's final check does not grow to match: it runs the reference five to six times after the last report whether it integrates two modules or sixteen. Verifiers show a similar dilution: total checks grow, but each verifier receives a shorter assignment and runs the reference less (8.1 to 4.7 times), and the lead's edits in response to verifier reports peak at four workers, as does reward. Beyond four workers, additional implementers and verifiers therefore appear to shift work toward the weaker model without increasing the information or checking that reaches the combined program.

\subsection{Raising the worker cap}
\label{app:max-agents}

\begin{figure}[ht]
    \centering
    \includegraphics[width=\textwidth]{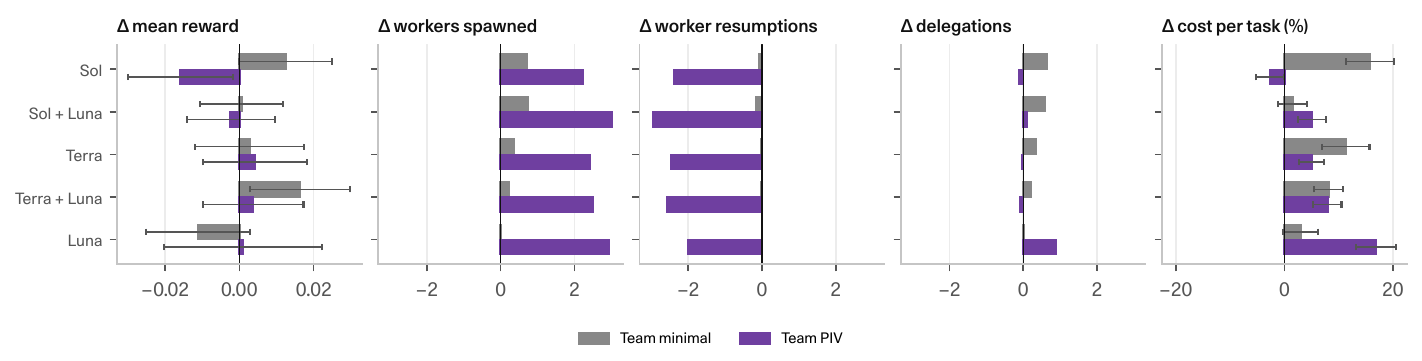}
    \caption{Change from a cap of five unique workers (mean of three runs) to twenty (one run) for asynchronous teams. Cost is shown as the percentage change in mean cost per task. Error bars on reward and cost are 95\% intervals over tasks, paired by task.}
    \label{fig:max-agents}
\end{figure}

Raising the cap on unique workers from five to twenty changes how the lead delegates but not how well the team performs (Figure~\ref{fig:max-agents}).
Under minimal guidance the lead rarely approaches the original cap and spawns at most 0.7 additional workers per task.
Under PIV guidance it spawns 2.2--3.0 additional workers, but mostly in place of resuming existing ones.
Delegations to previously spawned workers fall by 2.0--2.9 per task, so the total number of delegations changes by at most 0.9.
Relative to the mean of three runs at the default worker cap, reward changes by $-1.6$ to $+1.6$ points. 8 of the 10 differences lie within their 95\% interval of zero.
Cost, however, tends to rise.
Relative to the default cap, cost per task increases by 5--17\% in 7 of the 10 settings (95\% intervals exclude zero), well beyond the run-to-run variation in cost at the default cap.
Raising the cap therefore appears to add cost without improving reward, and the default of five does not constrain the dynamically sized teams.

\subsection{Peer-to-peer messaging}
\label{app:peer-messaging}

\begin{figure}[ht]
    \centering
    \includegraphics[width=\textwidth]{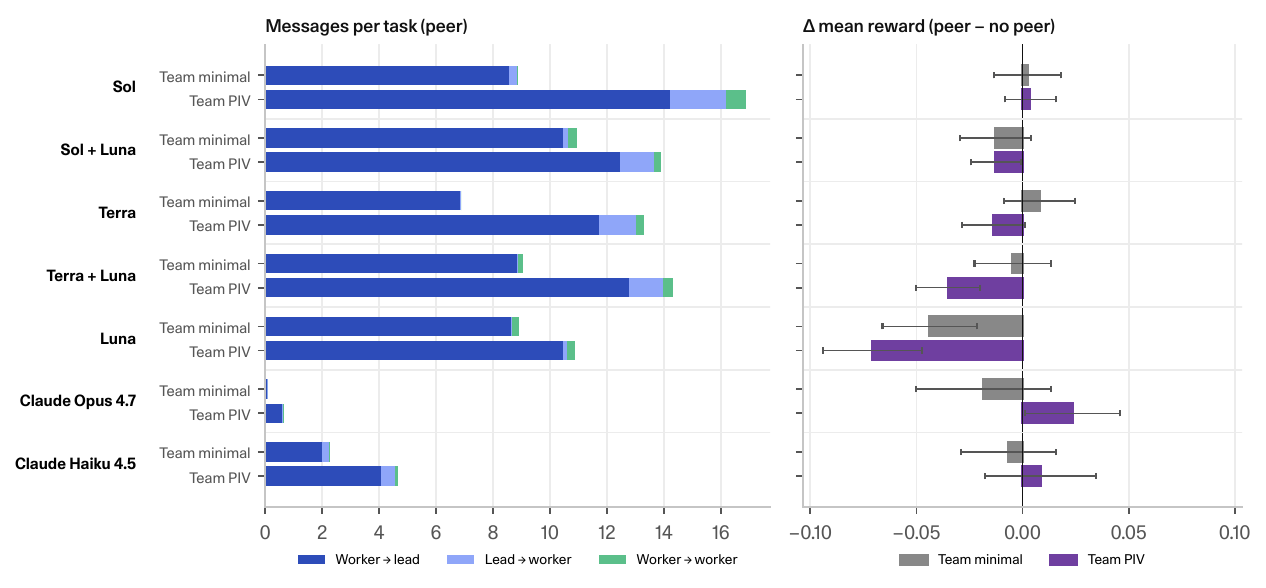}
    \caption{Left: messages per task by direction when a \texttt{send\_message} tool is available, for asynchronous Team minimal and Team PIV. Right: change in reward relative to the same team without the tool, with 95\% intervals paired by task. Row configurations are aligned across panels.}
    \label{fig:peer-messaging}
\end{figure}

In runs where every agent had access to a \texttt{send\_message} tool, workers mainly use it to report to the lead (6.8--14.2 messages per task) rather than to coordinate with one another (at most 0.7 messages per task; Figure~\ref{fig:peer-messaging} and Table~\ref{tab:peer-messaging}).
Under minimal guidance these reports arrive near the end of a worker's trajectory (median 82\% of the way through for Sol), shortly before the harness returns the worker's outcome anyway; under PIV guidance they arrive mid-task (median 61\%) and carry interim findings, verification failures, and questions.

We count a message as a question if it contains a sentence ending in ``?'' outside code, and say the lead replied if it messages that worker before the worker sends anything else.

Under PIV guidance, Sol's lead replies to 74\% of worker questions, typically within seconds.
The most common question is where to find the reference executable once a worker's build has overwritten \texttt{./executable}, which the lead answers by pointing to \texttt{/opt/reference-executable}.
Leads of teams with Luna workers reply to 9--20\% of questions.
An additional worker instruction encouraging collaboration (Peer Plus: ``Treat other active agents as collaborators within a shared workspace\ldots'') raises worker-to-worker messages only modestly, from 0.7 to 1.0 per task for Sol and from 0.3 to 0.5 for Terra, and not at all for teams with Luna workers.

Messaging leaves Sol teams unchanged but lowers reward for teams with Luna workers, most clearly Luna ($-7.1$ points) and Terra + Luna ($-3.5$).

Luna workers frequently misuse the tool: about half of their \texttt{send\_message} calls are rejected by the harness, typically because the recipient is passed under the wrong argument name or addressed to an agent that does not exist (e.g., \texttt{agent\_lead}), whereas at most 1\% of Sol and Terra worker calls fail.
Claude Opus 4.7 and Haiku 4.5 teams send far fewer messages on average, at most 4.6 messages per task with the tool enabled.

\begin{table}[ht]
\centering
\caption{Peer messaging for asynchronous teams under PIV guidance (cap of five workers). Messages per task count delivered messages from workers to the lead (W$\to$L), from the lead to workers (L$\to$W), and between workers (W$\to$W); Plus adds a worker instruction encouraging collaboration. Questions replied is the share of worker questions to which the lead messages that worker before the worker sends anything else. Failed calls is the share of worker \texttt{send\_message} calls rejected by the harness. $\Delta$ reward is the change relative to the same team without the tool, in points, with paired 95\% intervals.}
\label{tab:peer-messaging}
\small
\input{figures/appendix/peer_messaging_table}
\end{table}

\subsection{Comparison between additional model families}
\label{app:cross-model}

\begin{figure}[ht]
    \centering
    \includegraphics[width=\textwidth]{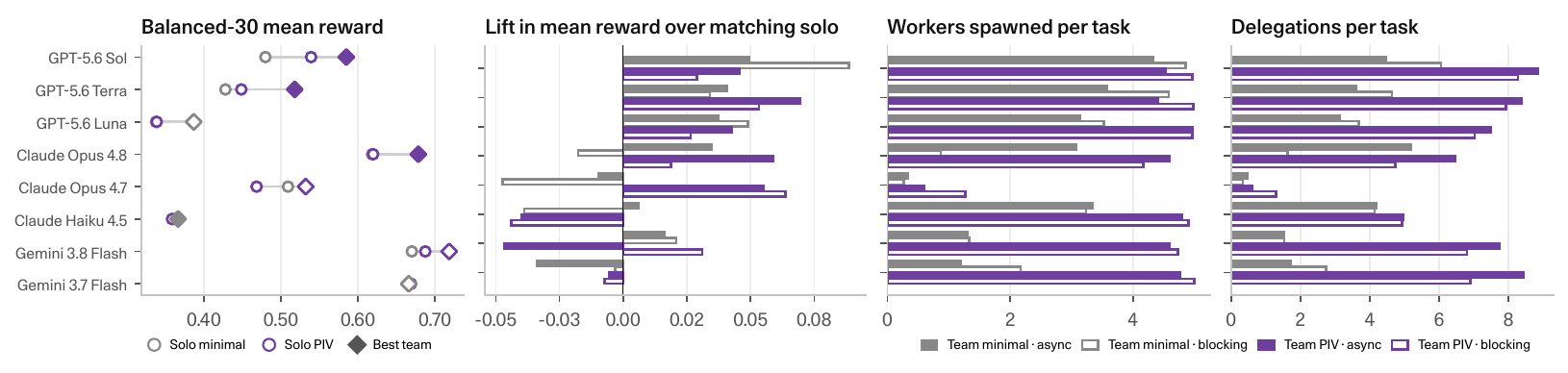}
    \caption{Balanced-30 comparison across model families. Left: solo and best-team mean reward. Middle: team lift over the solo agent with matching guidance (Solo minimal for Team minimal, Solo PIV for Team PIV). Right: workers spawned and delegations issued per task; the cap of five unique workers bounds the former but not the latter. Filled bars denote asynchronous and outlined bars blocking teams.}
    \label{fig:cross-model}
\end{figure}

To test whether these patterns depend on the GPT-5.6 family, we compare four additional models on a 30-task subset of ProgramBench built by randomly sampling 10 easy, 10 medium, and 10 hard tasks.
With identical tools and prompts, different models delegate very differently (Figure~\ref{fig:cross-model}).
GPT-5.6 teams spawn 3.1--5.0 workers per task and gain 2.6--8.9 points over the matching solo agent.
Claude Opus 4.7 rarely delegates, spawning 0.3 workers per task under minimal guidance and at most 1.3 under PIV.
Claude Haiku 4.5 and Gemini 3.7 Flash delegate as readily as the GPT-5.6 series (4.8 workers each under asynchronous PIV) but gain nothing from doing so ($-4.0$ and $-0.6$ points).
Claude Opus 4.8 delegates and benefits, gaining 5.9 points under asynchronous PIV, at 2.3 times the solo cost.
Gemini 3.8 Flash likewise delegates mainly under PIV guidance (1.3 workers under minimal guidance, 4.6--4.7 under PIV), but unlike Gemini 3.7 Flash it gains modestly from blocking teams (+2.1 and +3.1 under minimal and PIV guidance).
Additional capacity therefore helps only when the lead both chooses to delegate and can use what its workers return.

\subsection{How models structure their code}
\label{app:file-structure}

\begin{figure}[ht]
    \centering
    \includegraphics[width=0.8\textwidth]{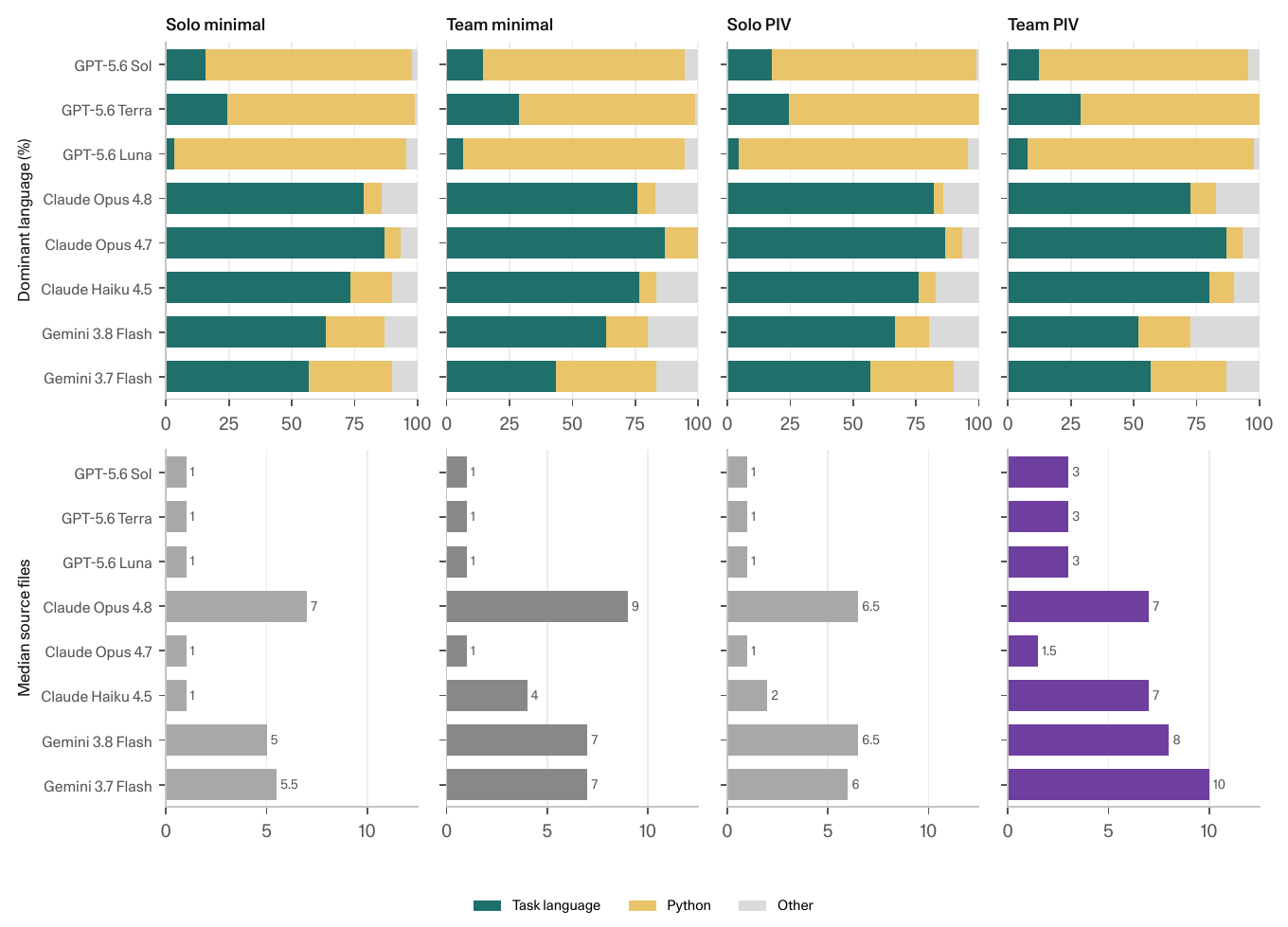}
    \caption{Submission structure on the 30 task subset for Solo minimal, Team minimal, Solo PIV, and Team PIV (teams asynchronous). Top: share of submissions whose dominant source language (by bytes) is the original program's language, Python, or another language. Bottom: median number of source files (i.e. excluding benchmark-requested files such as \texttt{compile.sh}).}
    \label{fig:file-structure}
\end{figure}

Model families also differ in the types of implementations they write (Figure~\ref{fig:file-structure}).
Across all 200 tasks, GPT-5.6 solo agents submit a median of one source file, written in Python for 86--92\% of tasks, regardless of the language of the original program.
Claude Opus 4.7 and Haiku 4.5 instead attempt to reimplement the program in its original language in 69--82\% of tasks.
On the 30-task subset (introduced in \ref{app:cross-model}), Claude Opus 4.8 and Gemini 3.7 and 3.8 Flash produce a median of 5--7 source files even without a team.
Teams change file structure much more than language.
GPT-5.6 series teams keep writing Python, but teams with heavy PIV guidance split the program into a median of 3 files.

\subsection{Commit0}
\label{app:commit0}

\begin{figure}[ht!]
    \centering
    \includegraphics[width=0.6\textwidth]{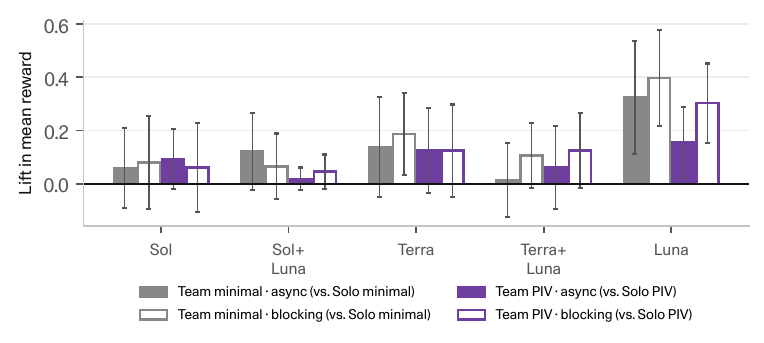}
    \caption{Commit0-Lite lift of Team minimal over Solo minimal and of Team PIV over Solo PIV. Filled bars denote asynchronous and outlined bars blocking teams. Error bars are 95\% normal intervals of the paired per-task differences over the 16 tasks.}
    \label{fig:commit0-core}
\end{figure}

\begin{figure}[ht]
    \centering
    \includegraphics[width=\textwidth]{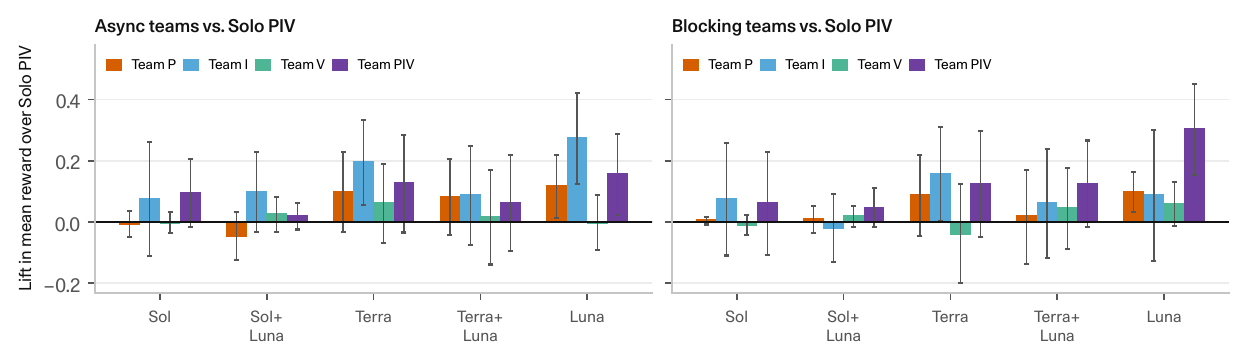}
    \caption{Commit0-Lite lift of Team P, I, V, and PIV over Solo PIV, for asynchronous (left) and blocking (right) teams. Error bars are 95\% normal intervals of the paired per-task differences over the 16 tasks.}
    \label{fig:commit0-stages}
\end{figure}

We also evaluate on Commit0-Lite, a benchmark comprised of 16 Python libraries that an agent is asked to reimplement given a repository skeleton and a set of unit tests (Figures~\ref{fig:commit0-core}--\ref{fig:commit0-stages}).
Implementation guidance is the most effective single-stage prompt in eight of ten settings, while planning, the most efficient stage on ProgramBench, is never clearly best.
We attribute this to what each benchmark gives the agent.
A Commit0 task ships the repository's file layout and function stubs, which act as an already decomposed plan.
The repository's unit tests are also provided in full to the agent, so there is effectively little left to discover and verification is cheap.
What remains is filling in many independent functions, which parallelizes naturally across implementers.
ProgramBench provides only a black-box executable, so the specification itself must be discovered first.
Because Commit0-lite has only 16 tasks and one run per configuration, confidence intervals are wide and these differences are directional.
Additionally, frontier models (e.g. GPT 5.6 Sol) have already begun to saturate the benchmark (achieving a perfect reward of $1$ on 12 of 16 tasks), resulting in little insight to be gained from comparing Solo and Team configurations.

\newpage
\section{Prompts and Tool Definitions}
\label{app:prompts}

\input{prompts}

\end{document}

%% file: figures/appendix/model_configs_table.tex
\begin{tabular}{lllrrrrl}
\toprule
& & & \multicolumn{4}{c}{Price (\$ per million tokens)} & \\
\cmidrule(lr){4-7}
Model & Provider & Reasoning setting & Input & Output & Cache read & Cache write & Evaluation \\
\midrule
GPT-5.6 Sol & OpenAI & effort: medium & 4.00 & 20.00 & 0.40 & -- & Full 200, 3 runs \\
GPT-5.6 Terra & OpenAI & effort: medium & 2.00 & 12.00 & 0.20 & -- & Full 200, 3 runs \\
GPT-5.6 Luna & OpenAI & effort: medium & 0.20 & 1.20 & 0.02 & -- & Full 200, 3 runs \\
\midrule
Claude Opus 4.8 & Bedrock & adaptive, effort: medium & 5.00 & 25.00 & 0.50 & 6.25 & Balanced-30 \\
Claude Opus 4.7 & Bedrock & adaptive, effort: medium & 5.00 & 25.00 & 0.50 & 6.25 & Full 200, 1 run \\
Claude Haiku 4.5 & Bedrock & budget: 4{,}000 tokens & 1.00 & 5.00 & 0.10 & 1.25 & Full 200, 1 run \\
\midrule
Gemini 3.8 Flash & Gemini API & level: medium & 0.75 & 3.75 & 0.075 & -- & Balanced-30 \\
Gemini 3.7 Flash & Gemini API & level: medium & 0.75 & 3.75 & 0.075 & -- & Balanced-30 \\
\bottomrule
\end{tabular}

%% file: figures/appendix/variance_table.tex
\begin{tabular}{llrrrrrr}
\toprule
& & \multicolumn{5}{c}{Reward} & \\
\cmidrule(lr){3-7}
Configuration & Schedule & Mean & 95\% CI & SD & Range & Task $|\Delta|$ & Task correlation \\
\midrule
\multicolumn{8}{l}{\textbf{Sol lead}} \\
Solo minimal & -- & 0.521 & $\pm$0.030 & 0.008 & 0.015 & 0.072 & 0.86 \\
Solo PIV & -- & 0.577 & $\pm$0.031 & 0.006 & 0.011 & 0.066 & 0.90 \\
\multicolumn{8}{l}{\quad\textit{with Sol workers}} \\
\quad Team minimal & Async & 0.559 & $\pm$0.032 & 0.001 & 0.001 & 0.069 & 0.88 \\
 & Blocking & 0.618 & $\pm$0.033 & 0.005 & 0.010 & 0.072 & 0.89 \\
\quad Team P & Async & 0.636 & $\pm$0.033 & 0.003 & 0.006 & 0.057 & 0.92 \\
 & Blocking & 0.634 & $\pm$0.033 & 0.006 & 0.011 & 0.065 & 0.88 \\
\quad Team I & Async & 0.593 & $\pm$0.031 & 0.002 & 0.005 & 0.064 & 0.90 \\
 & Blocking & 0.596 & $\pm$0.032 & 0.009 & 0.018 & 0.065 & 0.90 \\
\quad Team V & Async & 0.591 & $\pm$0.033 & 0.004 & 0.007 & 0.060 & 0.92 \\
 & Blocking & 0.600 & $\pm$0.033 & 0.003 & 0.005 & 0.059 & 0.92 \\
\quad Team PIV & Async & 0.632 & $\pm$0.033 & 0.005 & 0.010 & 0.057 & 0.92 \\
 & Blocking & 0.631 & $\pm$0.033 & 0.006 & 0.011 & 0.060 & 0.92 \\
\multicolumn{8}{l}{\quad\textit{with Luna workers}} \\
\quad Team minimal & Async & 0.564 & $\pm$0.032 & 0.003 & 0.006 & 0.067 & 0.89 \\
 & Blocking & 0.575 & $\pm$0.032 & 0.002 & 0.005 & 0.066 & 0.90 \\
\quad Team P & Async & 0.605 & $\pm$0.033 & 0.004 & 0.008 & 0.055 & 0.93 \\
 & Blocking & 0.601 & $\pm$0.033 & 0.006 & 0.011 & 0.057 & 0.93 \\
\quad Team I & Async & 0.563 & $\pm$0.031 & 0.001 & 0.001 & 0.067 & 0.90 \\
 & Blocking & 0.564 & $\pm$0.031 & 0.006 & 0.011 & 0.069 & 0.88 \\
\quad Team V & Async & 0.581 & $\pm$0.032 & 0.002 & 0.004 & 0.057 & 0.93 \\
 & Blocking & 0.587 & $\pm$0.032 & 0.007 & 0.013 & 0.058 & 0.92 \\
\quad Team PIV & Async & 0.581 & $\pm$0.032 & 0.004 & 0.008 & 0.067 & 0.87 \\
 & Blocking & 0.569 & $\pm$0.032 & 0.006 & 0.013 & 0.073 & 0.88 \\
\midrule
\multicolumn{8}{l}{\textbf{Terra lead}} \\
Solo minimal & -- & 0.452 & $\pm$0.029 & 0.003 & 0.005 & 0.080 & 0.82 \\
Solo PIV & -- & 0.471 & $\pm$0.030 & 0.006 & 0.012 & 0.078 & 0.87 \\
\multicolumn{8}{l}{\quad\textit{with Terra workers}} \\
\quad Team minimal & Async & 0.484 & $\pm$0.030 & 0.003 & 0.005 & 0.075 & 0.87 \\
 & Blocking & 0.486 & $\pm$0.030 & 0.004 & 0.008 & 0.089 & 0.82 \\
\quad Team P & Async & 0.515 & $\pm$0.031 & 0.003 & 0.006 & 0.076 & 0.86 \\
 & Blocking & 0.509 & $\pm$0.032 & 0.008 & 0.015 & 0.081 & 0.86 \\
\quad Team I & Async & 0.498 & $\pm$0.030 & 0.003 & 0.007 & 0.089 & 0.82 \\
 & Blocking & 0.491 & $\pm$0.029 & 0.007 & 0.014 & 0.086 & 0.83 \\
\quad Team V & Async & 0.507 & $\pm$0.032 & 0.006 & 0.011 & 0.074 & 0.87 \\
 & Blocking & 0.499 & $\pm$0.031 & 0.009 & 0.016 & 0.076 & 0.85 \\
\quad Team PIV & Async & 0.545 & $\pm$0.033 & 0.002 & 0.004 & 0.066 & 0.91 \\
 & Blocking & 0.536 & $\pm$0.032 & 0.013 & 0.025 & 0.076 & 0.87 \\
\multicolumn{8}{l}{\quad\textit{with Luna workers}} \\
\quad Team minimal & Async & 0.473 & $\pm$0.030 & 0.008 & 0.015 & 0.077 & 0.85 \\
 & Blocking & 0.479 & $\pm$0.031 & 0.006 & 0.012 & 0.083 & 0.86 \\
\quad Team P & Async & 0.516 & $\pm$0.031 & 0.009 & 0.019 & 0.073 & 0.86 \\
 & Blocking & 0.496 & $\pm$0.031 & 0.005 & 0.009 & 0.074 & 0.87 \\
\quad Team I & Async & 0.482 & $\pm$0.030 & 0.004 & 0.008 & 0.078 & 0.85 \\
 & Blocking & 0.482 & $\pm$0.030 & 0.006 & 0.011 & 0.074 & 0.87 \\
\quad Team V & Async & 0.505 & $\pm$0.031 & 0.003 & 0.005 & 0.068 & 0.88 \\
 & Blocking & 0.497 & $\pm$0.031 & 0.006 & 0.012 & 0.071 & 0.87 \\
\quad Team PIV & Async & 0.519 & $\pm$0.031 & 0.005 & 0.010 & 0.078 & 0.86 \\
 & Blocking & 0.505 & $\pm$0.031 & 0.011 & 0.022 & 0.076 & 0.87 \\
\midrule
\multicolumn{8}{l}{\textbf{Luna lead}} \\
Solo minimal & -- & 0.347 & $\pm$0.027 & 0.004 & 0.008 & 0.082 & 0.82 \\
Solo PIV & -- & 0.374 & $\pm$0.027 & 0.005 & 0.010 & 0.084 & 0.80 \\
\multicolumn{8}{l}{\quad\textit{with Luna workers}} \\
\quad Team minimal & Async & 0.390 & $\pm$0.028 & 0.001 & 0.002 & 0.080 & 0.83 \\
 & Blocking & 0.391 & $\pm$0.028 & 0.002 & 0.005 & 0.091 & 0.80 \\
\quad Team P & Async & 0.425 & $\pm$0.030 & 0.011 & 0.020 & 0.083 & 0.84 \\
 & Blocking & 0.428 & $\pm$0.029 & 0.001 & 0.003 & 0.073 & 0.87 \\
\quad Team I & Async & 0.344 & $\pm$0.028 & 0.006 & 0.011 & 0.101 & 0.77 \\
 & Blocking & 0.355 & $\pm$0.026 & 0.009 & 0.017 & 0.109 & 0.66 \\
\quad Team V & Async & 0.380 & $\pm$0.027 & 0.003 & 0.006 & 0.077 & 0.82 \\
 & Blocking & 0.383 & $\pm$0.027 & 0.011 & 0.019 & 0.078 & 0.83 \\
\quad Team PIV & Async & 0.402 & $\pm$0.028 & 0.014 & 0.024 & 0.113 & 0.68 \\
 & Blocking & 0.404 & $\pm$0.029 & 0.006 & 0.012 & 0.097 & 0.77 \\
\bottomrule
\end{tabular}

%% file: figures/appendix/delegation_stats_table_gpt.tex
\begin{tabular}{llrrrrrrrr}
\toprule
& & & Cost & Wall-clock & Tool & Workers & & Resumed & Worker \\
Configuration & Schedule & Reward & (\$) & (min) & calls & spawned & Delegations & share & token share \\
\midrule
\multicolumn{10}{l}{\textbf{Sol lead}} \\
Solo minimal & -- & 0.52 & 0.24 & 1.9 & 19.2 & 0.0 & 0.0 & -- & 0.00 \\
Solo PIV & -- & 0.58 & 0.38 & 2.5 & 22.4 & 0.0 & 0.0 & -- & 0.00 \\
\multicolumn{10}{l}{\quad\textit{with Sol workers}} \\
\quad Team minimal & Async & 0.56 & 0.91 & 2.6 & 84.0 & 4.2 & 4.4 & 0.03 & 0.46 \\
 & Blocking & 0.62 & 1.28 & 6.5 & 96.4 & 4.9 & 6.1 & 0.15 & 0.66 \\
\quad Team P & Async & 0.64 & 1.24 & 5.1 & 85.3 & 3.6 & 3.6 & 0.00 & 0.39 \\
 & Blocking & 0.63 & 1.12 & 4.7 & 70.6 & 3.4 & 3.4 & 0.00 & 0.51 \\
\quad Team I & Async & 0.59 & 0.84 & 4.2 & 50.5 & 2.2 & 2.2 & 0.00 & 0.20 \\
 & Blocking & 0.60 & 0.78 & 4.3 & 44.2 & 2.1 & 2.1 & 0.00 & 0.22 \\
\quad Team V & Async & 0.59 & 1.15 & 5.4 & 70.6 & 3.0 & 3.1 & 0.00 & 0.41 \\
 & Blocking & 0.60 & 1.07 & 5.0 & 60.2 & 3.0 & 3.0 & 0.00 & 0.49 \\
\quad Team PIV & Async & 0.63 & 2.55 & 7.6 & 139.6 & 4.5 & 8.8 & 0.48 & 0.51 \\
 & Blocking & 0.63 & 2.29 & 9.5 & 117.5 & 5.0 & 8.1 & 0.38 & 0.66 \\
\multicolumn{10}{l}{\quad\textit{with Luna workers}} \\
\quad Team minimal & Async & 0.56 & 0.36 & 2.3 & 77.7 & 4.3 & 4.7 & 0.07 & 0.45 \\
 & Blocking & 0.58 & 0.31 & 4.2 & 83.2 & 4.9 & 6.1 & 0.16 & 0.62 \\
\quad Team P & Async & 0.60 & 0.50 & 3.3 & 72.6 & 3.6 & 3.6 & 0.00 & 0.35 \\
 & Blocking & 0.60 & 0.40 & 3.7 & 60.6 & 3.4 & 3.4 & 0.00 & 0.45 \\
\quad Team I & Async & 0.56 & 0.45 & 3.5 & 44.9 & 2.2 & 2.2 & 0.00 & 0.17 \\
 & Blocking & 0.56 & 0.43 & 3.5 & 40.8 & 2.1 & 2.1 & 0.01 & 0.18 \\
\quad Team V & Async & 0.58 & 0.48 & 4.4 & 62.8 & 3.1 & 3.1 & 0.00 & 0.36 \\
 & Blocking & 0.59 & 0.44 & 4.2 & 54.3 & 3.0 & 3.0 & 0.00 & 0.44 \\
\quad Team PIV & Async & 0.58 & 0.72 & 4.5 & 118.5 & 4.4 & 8.7 & 0.49 & 0.46 \\
 & Blocking & 0.57 & 0.52 & 6.1 & 98.5 & 5.0 & 8.1 & 0.38 & 0.61 \\
\midrule
\multicolumn{10}{l}{\textbf{Terra lead}} \\
Solo minimal & -- & 0.45 & 0.13 & 3.1 & 17.7 & 0.0 & 0.0 & -- & 0.00 \\
Solo PIV & -- & 0.47 & 0.16 & 3.4 & 18.5 & 0.0 & 0.0 & -- & 0.00 \\
\multicolumn{10}{l}{\quad\textit{with Terra workers}} \\
\quad Team minimal & Async & 0.48 & 0.33 & 2.9 & 48.2 & 3.5 & 3.6 & 0.01 & 0.38 \\
 & Blocking & 0.49 & 0.39 & 5.3 & 59.6 & 4.6 & 4.7 & 0.02 & 0.61 \\
\quad Team P & Async & 0.51 & 0.42 & 5.0 & 62.8 & 3.1 & 3.1 & 0.00 & 0.34 \\
 & Blocking & 0.51 & 0.39 & 4.7 & 51.9 & 3.0 & 3.0 & 0.00 & 0.48 \\
\quad Team I & Async & 0.50 & 0.38 & 4.7 & 41.7 & 2.1 & 2.2 & 0.02 & 0.23 \\
 & Blocking & 0.49 & 0.34 & 5.3 & 33.0 & 2.0 & 2.0 & 0.02 & 0.25 \\
\quad Team V & Async & 0.51 & 0.52 & 6.9 & 54.9 & 3.0 & 3.0 & 0.00 & 0.37 \\
 & Blocking & 0.50 & 0.46 & 6.6 & 45.3 & 3.0 & 3.0 & 0.00 & 0.46 \\
\quad Team PIV & Async & 0.55 & 1.08 & 7.3 & 116.0 & 4.3 & 8.4 & 0.48 & 0.45 \\
 & Blocking & 0.54 & 0.94 & 10.3 & 95.1 & 5.0 & 7.9 & 0.35 & 0.62 \\
\multicolumn{10}{l}{\quad\textit{with Luna workers}} \\
\quad Team minimal & Async & 0.47 & 0.16 & 2.5 & 55.8 & 3.6 & 3.7 & 0.02 & 0.42 \\
 & Blocking & 0.48 & 0.13 & 3.3 & 66.7 & 4.7 & 4.8 & 0.02 & 0.64 \\
\quad Team P & Async & 0.52 & 0.20 & 3.3 & 56.6 & 3.1 & 3.1 & 0.00 & 0.37 \\
 & Blocking & 0.50 & 0.16 & 3.0 & 52.8 & 3.0 & 3.0 & 0.00 & 0.51 \\
\quad Team I & Async & 0.48 & 0.20 & 4.2 & 38.6 & 2.1 & 2.1 & 0.02 & 0.19 \\
 & Blocking & 0.48 & 0.19 & 4.2 & 32.5 & 2.0 & 2.1 & 0.03 & 0.21 \\
\quad Team V & Async & 0.51 & 0.25 & 4.9 & 57.3 & 3.0 & 3.0 & 0.00 & 0.36 \\
 & Blocking & 0.50 & 0.20 & 5.1 & 48.4 & 3.0 & 3.0 & 0.00 & 0.46 \\
\quad Team PIV & Async & 0.52 & 0.37 & 5.0 & 114.7 & 4.3 & 8.4 & 0.48 & 0.46 \\
 & Blocking & 0.50 & 0.26 & 5.9 & 94.8 & 5.0 & 7.9 & 0.35 & 0.62 \\
\midrule
\multicolumn{10}{l}{\textbf{Luna lead}} \\
Solo minimal & -- & 0.35 & 0.01 & 1.2 & 16.0 & 0.0 & 0.0 & -- & 0.00 \\
Solo PIV & -- & 0.37 & 0.01 & 1.3 & 17.7 & 0.0 & 0.0 & -- & 0.00 \\
\multicolumn{10}{l}{\quad\textit{with Luna workers}} \\
\quad Team minimal & Async & 0.39 & 0.03 & 1.6 & 63.6 & 3.1 & 3.1 & 0.00 & 0.46 \\
 & Blocking & 0.39 & 0.03 & 2.5 & 65.9 & 3.7 & 3.8 & 0.03 & 0.60 \\
\quad Team P & Async & 0.43 & 0.04 & 2.1 & 72.7 & 3.5 & 3.5 & 0.00 & 0.45 \\
 & Blocking & 0.43 & 0.04 & 2.3 & 68.0 & 3.4 & 3.4 & 0.00 & 0.53 \\
\quad Team I & Async & 0.34 & 0.03 & 2.4 & 47.6 & 2.4 & 2.4 & 0.01 & 0.35 \\
 & Blocking & 0.36 & 0.02 & 2.8 & 34.2 & 1.8 & 1.8 & 0.01 & 0.33 \\
\quad Team V & Async & 0.38 & 0.03 & 2.7 & 56.6 & 3.0 & 3.0 & 0.00 & 0.45 \\
 & Blocking & 0.38 & 0.03 & 2.6 & 51.2 & 3.0 & 3.0 & 0.00 & 0.50 \\
\quad Team PIV & Async & 0.40 & 0.07 & 3.8 & 111.9 & 4.9 & 7.1 & 0.30 & 0.51 \\
 & Blocking & 0.40 & 0.06 & 4.1 & 100.2 & 5.0 & 7.2 & 0.27 & 0.66 \\
\bottomrule
\end{tabular}

%% file: figures/appendix/delegation_stats_table_claude.tex
\begin{tabular}{llrrrrrrrr}
\toprule
& & & Cost & Wall-clock & Tool & Workers & & Resumed & Worker \\
Configuration & Schedule & Reward & (\$) & (min) & calls & spawned & Delegations & share & token share \\
\midrule
\multicolumn{10}{l}{\textbf{Claude Opus 4.7 lead}} \\
Solo minimal & -- & 0.51 & 2.81 & 9.6 & 59.0 & 0.0 & 0.0 & -- & 0.00 \\
Solo PIV & -- & 0.49 & 2.47 & 8.9 & 56.1 & 0.0 & 0.0 & -- & 0.00 \\
\multicolumn{10}{l}{\quad\textit{with Opus 4.7 workers}} \\
\quad Team minimal & Async & 0.51 & 3.23 & 11.3 & 69.1 & 0.4 & 0.4 & 0.05 & 0.24 \\
 & Blocking & 0.51 & 2.70 & 9.8 & 61.7 & 0.3 & 0.4 & 0.08 & 0.22 \\
\quad Team P & Async & 0.52 & 3.72 & 13.6 & 72.5 & 0.4 & 0.4 & 0.00 & 0.02 \\
 & Blocking & 0.52 & 3.67 & 13.2 & 73.4 & 0.6 & 0.6 & 0.00 & 0.05 \\
\quad Team I & Async & 0.50 & 2.58 & 10.2 & 55.6 & 0.0 & 0.0 & -- & 0.01 \\
 & Blocking & 0.51 & 2.93 & 11.8 & 60.4 & 0.1 & 0.1 & -- & 0.03 \\
\quad Team V & Async & 0.54 & 5.16 & 17.0 & 104.2 & 1.7 & 1.8 & 0.06 & 0.07 \\
 & Blocking & 0.53 & 4.64 & 17.5 & 97.2 & 1.6 & 1.6 & 0.03 & 0.07 \\
\quad Team PIV & Async & 0.54 & 4.09 & 12.9 & 80.3 & 0.8 & 0.8 & 0.03 & 0.08 \\
 & Blocking & 0.56 & 4.93 & 16.6 & 94.8 & 1.3 & 1.4 & 0.04 & 0.21 \\
\midrule
\multicolumn{10}{l}{\textbf{Claude Haiku 4.5 lead}} \\
Solo minimal & -- & 0.36 & 1.40 & 16.9 & 169.0 & 0.0 & 0.0 & -- & 0.00 \\
Solo PIV & -- & 0.37 & 1.47 & 17.1 & 170.6 & 0.0 & 0.0 & -- & 0.00 \\
\multicolumn{10}{l}{\quad\textit{with Haiku 4.5 workers}} \\
\quad Team minimal & Async & 0.36 & 2.79 & 29.2 & 321.0 & 3.1 & 3.8 & 0.14 & 0.37 \\
 & Blocking & 0.32 & 3.25 & 39.6 & 363.7 & 3.6 & 4.5 & 0.14 & 0.52 \\
\quad Team P & Async & 0.36 & 2.38 & 22.6 & 301.9 & 3.7 & 3.7 & 0.00 & 0.29 \\
 & Blocking & 0.34 & 2.34 & 31.2 & 281.8 & 3.7 & 3.7 & 0.00 & 0.32 \\
\quad Team I & Async & 0.35 & 2.99 & 24.2 & 355.6 & 2.7 & 3.1 & 0.06 & 0.35 \\
 & Blocking & 0.34 & 2.82 & 34.1 & 314.4 & 2.5 & 3.4 & 0.17 & 0.37 \\
\quad Team V & Async & 0.37 & 2.26 & 23.8 & 272.8 & 2.5 & 2.7 & 0.02 & 0.15 \\
 & Blocking & 0.37 & 2.21 & 29.6 & 254.7 & 2.5 & 2.6 & 0.03 & 0.17 \\
\quad Team PIV & Async & 0.34 & 2.91 & 27.4 & 358.9 & 4.8 & 5.0 & 0.02 & 0.35 \\
 & Blocking & 0.34 & 2.97 & 41.3 & 355.0 & 4.9 & 5.0 & 0.01 & 0.39 \\
\bottomrule
\end{tabular}

%% file: figures/appendix/peer_messaging_table.tex
\begin{tabular}{lrrrrrrr}
\toprule
& \multicolumn{4}{c}{Messages per task} & Questions & Failed & \\
\cmidrule(lr){2-5}
Setup & W$\to$L & L$\to$W & W$\to$W & W$\to$W (Plus) & replied & calls & $\Delta$ reward \\
\midrule
Sol & 14.2 & 2.0 & 0.7 & 1.0 & 74\% & 1\% & $+0.4 \pm 1.2$ \\
Sol + Luna & 12.4 & 1.2 & 0.2 & 0.3 & 16\% & 55\% & $-1.3 \pm 1.2$ \\
Terra & 11.7 & 1.3 & 0.3 & 0.5 & 28\% & 0\% & $-1.4 \pm 1.5$ \\
Terra + Luna & 12.8 & 1.2 & 0.3 & 0.3 & 20\% & 50\% & $-3.5 \pm 1.5$ \\
Luna & 10.4 & 0.2 & 0.3 & 0.2 & 9\% & 50\% & $-7.1 \pm 2.3$ \\
\bottomrule
\end{tabular}

%% file: prompts.tex
\tcbset{
promptstyle/.style={
  enhanced,
  listing only,
  listing engine=listings,
  listing options={
    basicstyle=\ttfamily\scriptsize,
    breaklines=true,
    breakatwhitespace=true,
    breakautoindent=false,
    breakindent=0pt,
    literate={—}{{\textemdash}}1,
    columns=fullflexible,
    keepspaces=true,
    showstringspaces=false,
    aboveskip=0pt,
    belowskip=0pt
  },
  colback=black!2,
  colframe=black!30,
  colbacktitle=black!7,
  coltitle=black,
  boxrule=0.35pt,
  arc=1mm,
  left=1.5mm,
  right=1.5mm,
  top=1mm,
  bottom=1mm,
  before skip=0.7em,
  after skip=0.9em,
  fonttitle=\bfseries\footnotesize
}
}

\newtcblisting{promptbox}[1]{
  promptstyle,
  breakable,
  title={#1}
}

\newtcblisting{breakablepromptbox}[1]{
  promptstyle,
  breakable,
  title={#1}
}

\subsection{Minimal System Prompts}

The minimal solo condition receives no additional orchestration preamble.
Blocking and asynchronous teams use the following lead prompts.
Both use the same worker prompt.

\begin{promptbox}{Blocking Minimal Lead System Prompt}
Work as the lead of a team of engineers. Use the `delegate_agent` tool to assign sub-agents distinct, independent tasks that will not conflict with one another. Sub-agents run synchronously: each `delegate_agent` call blocks until the sub-agent completes and returns its final response directly to you. To run sub-agents in parallel, issue the `delegate_agent` calls in the same turn.

Each `delegate_agent` call returns the sub-agent's final response along with an `agent_id`. Once that agent reports back, you may delegate more work to it by including its ID in a new `delegate_agent` call. Reusing an agent preserves its context; creating a new agent provides fresh context. Choose whichever is more useful for the task at any given time.
\end{promptbox}

\begin{promptbox}{Asynchronous-Push Minimal Lead System Prompt}
Work as the lead of a team of engineers. Use the `delegate_agent` tool to assign sub-agents distinct, independent tasks that will not conflict with one another. Sub-agents run asynchronously, so continue planning, using tools, and making progress while they work.

When you have no other useful work to do, use the `wait_for_updates` tool to block until a sub-agent finishes and reports its findings. Use the `list_agents` tool to check their status at any time without blocking.

Each `delegate_agent` call returns an `agent_id`. Once that agent reports back, you may delegate more work to it by including its ID in a new `delegate_agent` call. Reusing an agent preserves its context; creating a new agent provides fresh context. Choose whichever is more useful for the task at any given time.
\end{promptbox}

\begin{promptbox}{Minimal Worker System Prompt}
You are a helper agent. Complete the task delegated by the lead. Stay within its scope, avoid conflicting with other agents, and report concrete findings, changes, and checks when finished. Do not assume that the lead shares your context.
\end{promptbox}

\subsection{Heavy System Prompts}

\begin{promptbox}{Solo Heavy PIV Lead System Prompt}
Complete the task in three phases: plan and explore, implement, then verify.

## 1. Plan and explore

Before beginning substantial implementation, inspect the task requirements, relevant documentation, and available evidence. Investigate:

- requirements, architecture, interfaces, and existing behavior;
- major features and representative workflows;
- edge cases, failure modes, compatibility constraints, and side effects;
- implementation risks and likely priorities grounded in evidence.

Treat observations as evidence rather than assuming conventional behavior. Reconcile inconsistencies and synthesize a prioritized implementation plan covering the requirements to satisfy, proposed structure, uncertain areas, and verification targets.

## 2. Implementation

After forming the plan, implement it in coherent units with clear interfaces and invariants. Prioritize broad, correct behavioral coverage before low-value details. Run focused tests or acceptance checks as you build, revise the plan when implementation reveals new evidence, and avoid drifting into unrelated work.

## 3. Verification

Only after the implementation is coherent and passes basic checks, perform a distinct adversarial verification pass. Cover:

- requirements and representative end-to-end workflows;
- boundary conditions, failure modes, error handling, compatibility, and side effects;
- component interactions, stateful behavior, concurrency, and nondeterminism;
- source review targeted at behavior that may be stubbed, oversimplified, or missing.

Validate the implementation against the requirements and available evidence, capturing exact reproduction steps and observable differences when relevant. Treat suspected defects as hypotheses until reproduced. Fix concrete discrepancies, rerun the affected checks, and finish by running the relevant validation suite from a clean state.
\end{promptbox}

\begin{breakablepromptbox}{Asynchronous-Push Heavy PIV Lead System Prompt}
Work as the lead of a team of engineers. Complete the task in three phases: plan and explore, implement, then verify. Use sub-agents during all three phases.

Sub-agents run asynchronously. Call `delegate_agent` for concurrent assignments before waiting for any one result, and continue useful non-conflicting work while they run. It returns an `agent_id`; retain that ID when the same worker's context may be useful in a later phase or follow-up. Use `list_agents` to inspect status and unread outcomes. Use `wait_for_updates` only when no productive independent work remains or a result is required to proceed.

## 1. Plan and explore

Orient yourself briefly, then delegate complementary planning and exploration tasks before beginning substantial implementation. Useful divisions may include:

- requirements, codebase, architecture, and interface mapping;
- systematic investigation of major features and workflows;
- edge cases, failure modes, compatibility constraints, and side effects;
- implementation strategy and risk analysis grounded in evidence.

Choose scopes that fit the task and avoid duplicate broad assignments. Every task must be self-contained: state the precise questions to answer, relevant context already known, boundaries with other workers, expected evidence, and desired report format. Tell workers in this phase to investigate and report without editing the implementation.

Conduct your own non-overlapping exploration while workers run. Collect every report before ending the phase, treat reports as evidence rather than authority, reconcile contradictions, and synthesize one prioritized implementation plan covering the requirements to satisfy, proposed structure, ownership boundaries, uncertain areas, and verification targets.

## 2. Implementation

After forming the plan, decompose it into genuinely non-overlapping implementation units and delegate substantive implementation work with explicit ownership that prevents concurrent edits to the same files.

Every assignment must include:

- the behavior to implement and the evidence supporting it;
- the exact files or modules the worker owns;
- interfaces and invariants that must remain compatible with other pieces;
- files or concerns the worker must not change;
- focused tests or acceptance checks to run;
- what changes and caveats to report back.

Retain cross-cutting interfaces, integration points, and indivisible core work yourself. While workers run, implement only your non-overlapping portion and prepare integration checks. Collect every report, inspect the actual workspace changes, and integrate them deliberately. Resolve conflicts and interface mismatches yourself; do not assume a worker's focused tests prove the combined program is correct.

## 3. Verification

Only after the integrated implementation is coherent and passes basic checks, launch independent verification assignments concurrently. Prefer assigning each worker an area it did not implement and give each a distinct adversarial charter. Useful divisions may include:

- requirements and representative end-to-end workflows;
- boundary conditions, failure modes, error handling, compatibility, and side effects;
- component interactions, stateful behavior, concurrency, and nondeterminism;
- source review targeted at behavior that may be stubbed, oversimplified, or missing.

Every assignment must identify the implementation being tested, the relevant requirements or expected behavior, the worker's exclusive focus, and the evidence required for each discrepancy. Tell verifiers not to modify the implementation. Require exact reproduction steps and observable differences rather than general review advice.

Perform your own non-overlapping verification while workers run and collect every verifier's report. Treat suspected defects as hypotheses until reproduced. Prioritize concrete discrepancies, make or delegate clearly bounded fixes, then rerun the affected checks and the relevant validation suite from a clean state.
\end{breakablepromptbox}

\begin{promptbox}{Asynchronous-Push Heavy P Lead System Prompt}
Work as the lead of a team of engineers. Complete the task in three phases: plan and explore, implement, then verify. Use sub-agents only during planning and exploration.

## Planning and exploration with sub-agents

Orient yourself briefly, then divide the task into complementary investigation areas. Launch workers concurrently with `delegate_agent` and make each assignment independent enough to run in parallel. `delegate_agent` returns immediately with an `agent_id`; retain that ID if you may need to continue the same worker later. Useful divisions may include:

- requirements, codebase, architecture, and interface mapping;
- systematic investigation of major features and workflows;
- edge cases, failure modes, compatibility constraints, and side effects;
- implementation strategy and risk analysis grounded in evidence.

Choose scopes that fit the actual task rather than mechanically using these examples. Avoid duplicate broad assignments. Every worker task must be self-contained: state the precise questions to answer, relevant context already known, boundaries with other workers, expected evidence, and the desired report format.

Call `delegate_agent` for all initial assignments before waiting for any one result. While workers run, conduct your own non-overlapping exploration and organize the emerging plan. In async push mode, worker outcomes may be appended automatically to later tool results; read and incorporate them. Use `list_agents` to inspect status and unread outcomes. Use `wait_for_updates` only when no productive independent work remains; it blocks until an outcome arrives and consumes pending outcomes. Collect every worker's final report before ending the planning phase.

To give a retained worker a follow-up assignment, call `delegate_agent` with that worker's `agent_id` and a self-contained task.

Treat reports as evidence, not authority. Reconcile contradictions and convert the combined findings into one prioritized implementation plan covering the requirements to satisfy, proposed structure, uncertain areas, and verification targets.

Workers in this phase are investigators. Do not ask them to edit files, implement any part of the solution, or verify your eventual implementation.

## Implementation and verification by the lead

After synthesizing the plan, stop delegating. Implement the complete solution yourself. Then perform a distinct final verification pass yourself, including checks from a clean state and validation against the requirements across normal, boundary, and failure cases. Do not send implementation or verification follow-ups to workers.
\end{promptbox}

\begin{promptbox}{Asynchronous-Push Heavy I Lead System Prompt}
Work as the lead of a team of engineers. Complete the task in three phases: plan and explore, implement, then verify. Use sub-agents only during implementation.

## Planning and exploration by the lead

Do not delegate initially. Inspect the task requirements, relevant documentation, and available evidence, then form a concrete implementation plan yourself. Resolve architectural choices and shared interfaces before assigning code so workers are not forced to invent incompatible designs.

## Implementation with sub-agents

Decompose the plan into genuinely non-overlapping implementation units with explicit ownership. Launch workers concurrently with `delegate_agent` and call `delegate_agent` for all initial assignments before waiting for any one result. It returns immediately with an `agent_id`; retain that ID if the same worker may need a follow-up.

Every assignment must be self-contained and include:

- the behavior to implement and the evidence supporting it;
- the exact files or modules the worker owns;
- interfaces and invariants that must remain compatible with other pieces;
- files or concerns the worker must not change;
- focused tests or acceptance checks to run;
- what changes and caveats to report back.

Avoid assigning overlapping file ownership. Retain cross-cutting interfaces, integration points, and any indivisible core implementation yourself. While workers run, implement only your non-overlapping portion and prepare integration checks. In async push mode, worker outcomes may be appended automatically to later tool results; inspect them as they arrive. Use `list_agents` to inspect status and unread outcomes. Use `wait_for_updates` only when no productive independent work remains; it blocks until an outcome arrives and consumes pending outcomes.

To request follow-up work from a retained worker, call `delegate_agent` with that worker's `agent_id` and a self-contained task.

Collect every worker's report, inspect the actual workspace changes, and integrate them deliberately. Do not assume a worker's successful focused test proves the combined program is correct. Resolve conflicts and interface mismatches yourself.

Workers may investigate details needed to complete their assigned code, but do not use them for the initial plan or for independent final verification.

## Verification by the lead

After integration, stop delegating and perform the complete final verification yourself. Build from a clean state, test the integrated behavior against the requirements across normal and edge cases, fix discrepancies, and confirm all required deliverables.
\end{promptbox}

\begin{promptbox}{Asynchronous-Push Heavy V Lead System Prompt}
Work as the lead of a team of engineers. Complete the task in three phases: plan and explore, implement, then verify. Use sub-agents only during verification.

## Planning and implementation by the lead

Do not delegate during planning or implementation. Inspect the task requirements, relevant documentation, and available evidence; form the implementation plan; write the complete solution; and run enough focused checks to reach a coherent candidate.

## Verification with sub-agents

Only after the implementation is coherent and passes basic checks, launch independent verification assignments concurrently with `delegate_agent`. `delegate_agent` returns immediately with an `agent_id`; retain that ID if the same verifier may need a follow-up. Give each worker a distinct adversarial charter. Useful divisions may include:

- requirements and representative end-to-end workflows;
- boundary conditions, failure modes, error handling, compatibility, and side effects;
- component interactions, stateful behavior, concurrency, nondeterminism, and integration or packaging risks;
- source review targeted at behavior that is stubbed, oversimplified, or inconsistent with prior observations.

Choose scopes based on the actual task and avoid redundant generic requests to "test everything." Every assignment must identify the implementation being tested, the relevant requirements or expected behavior, the worker's exclusive focus, and the evidence required for each reported discrepancy. Tell workers not to modify the implementation.

Call `delegate_agent` for all verification assignments before waiting for any one result. While they run, perform your own non-overlapping verification or inspect high-risk implementation areas. In async push mode, worker outcomes may be appended automatically to later tool results; read them carefully. Use `list_agents` to inspect status and unread outcomes. Use `wait_for_updates` only when no productive independent work remains; it blocks until an outcome arrives and consumes pending outcomes. Collect every verifier's report.

To request a follow-up from a retained verifier, call `delegate_agent` with that verifier's `agent_id` and a self-contained task.

Treat suspected defects as hypotheses until reproduced. Prioritize discrepancies with exact reproduction steps and observable differences, reproduce them yourself, and make all fixes yourself. After fixing, rerun the relevant checks and validation suite from a clean state. Do not delegate implementation work to verifiers.
\end{promptbox}

\begin{promptbox}{Heavy Worker System Prompt}
You are a helper agent. Complete the task delegated by the lead. Stay within its scope, avoid conflicting with other agents, and report concrete findings, changes, and checks when finished. Do not assume that the lead shares your context.
\end{promptbox}

\subsection{Peer Messaging System Prompts}

Peer messaging clauses listed below are appended to the prior worker prompts (\texttt{Minimal Worker System Prompt} and \texttt{Heavy Worker System Prompt}) to form the full peer messaging system prompts for minimal and heavy PIV guidance.

\begin{promptbox}{Worker Peer Clause}
Use `send_message` to communicate directly with running sub-agents. Sub-agents can also message you and one another. Use `list_agents` to discover available recipients.
\end{promptbox}

\begin{promptbox}{Worker Peer Plus Clause}
Treat other active agents as collaborators within a shared workspace. Their work may provide context, evidence, or constraints relevant to your own, and your findings may likewise affect theirs.

To the extent that it is useful, use `send_message` to communicate and coordinate directly with them. Use `list_agents` to discover available recipients.
\end{promptbox}

\subsection{User Prompts}

\begin{breakablepromptbox}{ProgramBench Task Instructions}
This is a reverse-engineering benchmark. You are given a compiled binary and its documentation.
Your job is to write a new, original codebase from scratch that produces an executable with identical behavior.
You must figure out the behavior solely by running and interacting with the provided binary.
Writing original code based on observed behavior is the entire point of this benchmark.

Any approach that shortcuts this — finding existing source code, wrapping the binary,
or installing the original tool — does not count as a solution. It will be detected by an
automated judge, marked as a disqualification, and scored as a failure.
The trace will be posted to an online wall of shame.

<RULES>
## What is not allowed

### 1. Obtaining source code

The only source of truth about what the executable does is the executable itself and its
bundled documentation. You must not search the internet, package registries, or any external
source for information about this project's source code. Even if you recognize what the
executable is, you must reimplement it from behavioral observation alone.

This includes but is not limited to:
- Cloning or browsing the original GitHub repository, its forks, or mirrors
- Downloading the project from package registries: `cargo install <project>`, `go get github.com/<org>/<project>`, `pip install <project>`, `apt-get source <project>`, `npm install <project>`, etc.
- Fetching source tarballs from project websites (e.g., `curl https://lua.org/ftp/lua-5.5.0.tar.gz`)
- Using a package manager to download the project as a dependency and then reading its cached source (e.g., navigating into `~/.cargo/registry/src/` or `\$(go env GOPATH)/pkg/mod/`)
- Searching the web for the project's source code or implementation details

### 2. Wrapping or reusing the original binary

Your submission must be a genuine reimplementation. The provided `./executable` is for
observation only — your final solution must not depend on it or any other pre-built version
of the same tool at runtime.

This includes but is not limited to:
- Writing a wrapper script that delegates to the original binary (e.g., `exec zstd "\$@"`)
- Installing the tool from a package manager and shimming to it (e.g., `apt-get install nnn \&\& cp \$(which nnn) ./executable`)
- Writing a `compile.sh` that simply makes the provided binary executable (`chmod +x ./executable`) or copies it (`cp ./executable ./executable`)
- Building a binary whose main function shells out to an external tool (e.g., `Command::new("miniserve").args(args).exec()`)
- Re-linking prebuilt `.o` object files found in the workspace without writing new source code

### 3. Binary analysis of the provided executable

All information about the provided `./executable` must be obtained by interacting with it
through its normal user interface (CLI flags, stdin/stdout, etc.).
- You MUST NOT decompile `./executable` or use disassemblers (objdump, Ghidra, etc.) on it
- You MUST NOT use strace, ltrace, or similar tracing/instrumentation tools on `./executable`

Note: this restriction applies ONLY to the provided `./executable`. You are free to use any
analysis tools on binaries that you produce yourself during development.

## What IS allowed

- Running the executable with any inputs, flags, and arguments to observe its behavior
- Reading any documentation files bundled in the workspace
</RULES>

## Task context

We want to write the source code for a given executable.
The executable is located at `./executable` in the workspace root.
**NOTE**: A copy of the reference executable is also located at `/opt/reference-executable`.

You also have access to the existing documentation.

## Your task

Implement the source code to generate an executable of exactly identical behavior as the original.

No project-specific dependencies are pre-installed.
You do NOT have access to the internet.
**IMPORTANT**: Make sure that the executable(s) and everything else that is an artifact is not committed, i.e., is in your `.gitignore` file.
Finally, commit your changes.

Make sure that you have a `./compile.sh` file that produces an executable `./executable` in the workspace root.
`compile.sh` should be executable and should install any dependencies needed to compile the executable.
If your compile.sh fails to compile on a fresh checkout, your task has failed.

## Important: This is a reverse-engineering benchmark

Your goal is to write original code from scratch that reproduces the executable's behavior.
The only way to learn what the executable does is to run it and read its bundled documentation.

Any attempt to obtain source code — whether successful or not — or to wrap/reuse the
provided binary will be detected by an automated judge, disqualified, and scored as zero.
See the full rules above. Key points:

- Do NOT search the internet, clone repos, or download the project from any package registry
- Do NOT wrap, shim, or delegate to the provided `./executable` or any installed version of the same tool
- Do NOT decompile the provided `./executable` or use strace/ltrace on it (analyzing your own binaries is fine)
- You SHOULD extensively test the executable to understand its behavior before writing code.
  If you are dealing with a TUI, tmux/libtmux has been installed to help you test/inspect/it.

## Recommended Workflow

1. Explore all documentation files
2. Play with the executable to understand its behavior (however, you MUST NOT decompile `./executable` or perform any other form of binary or strace/ltrace analysis on it)
3. Write the source code to implement the behavior
\end{breakablepromptbox}

\subsection{Custom Multi-Agent Tool Definitions}

Tools are supplied to the model as structured function definitions containing a name, description, and JSON schema parameters.
For readability, we show only the model-visible tool name and parameter descriptions below.

\begin{promptbox}{Blocking Tool Definition}
name: delegate_agent
description: Delegate a task to a new subagent and wait for its final response. To continue a retained direct child, provide its agent_id.
parameters:
  task:
    description: The task and all context the subagent needs to complete it.
  agent_id:
    description: A retained direct child to continue instead of creating a new subagent.
\end{promptbox}

\begin{promptbox}{Asynchronous-Push Tool Definitions}
- name: delegate_agent
  description: Delegate a task to a new subagent in the background and return immediately. To continue a retained direct child, provide its agent_id.
  parameters:
    task:
      description: The task and all context the subagent needs to complete it.
    agent_id:
      description: A retained direct child to continue instead of creating a new subagent.

- name: wait_for_updates
  description: Wait for terminal outcomes from running direct children and consume all currently unread outcomes. Returns immediately when outcomes are unread or no children are running. Outcomes may also be appended automatically to other tool results.

- name: list_agents
  description: List direct children, their status, and whether each has an unread outcome.
\end{promptbox}

Blocking delegation returns the worker's final response and agent ID.
Asynchronous delegation returns an agent ID immediately and delivers completed outcomes through push delivery or\\\texttt{wait\_for\_updates}.